\documentclass[10pt,a4paper,twocolumn,multicol,superscriptaddress,prb,amsmath,amssymb,aps,prb]{revtex4}
\usepackage{graphicx}
\usepackage{bm}

\newcommand{\kk}{{\bf k}}

\def  \bsig    {\mbox{\boldmath$\sigma $}}

\begin{document}
\title{Anomalous Hall effect in 2D Dirac band: link between Kubo-Streda formula and semiclassical Boltzmann equation approach} 

\author{N.A. Sinitsyn}
\affiliation{Department of Physics, Texas A\&M University,
College Station, TX 77843-4242, USA}
\affiliation{Department of Physics, University of Texas at Austin,
Austin TX 78712-1081, USA}
\author{A.H. MacDonald} 
\affiliation{Department of Physics, University of Texas at Austin,
Austin TX 78712-1081, USA}
\author{T. Jungwirth}
\affiliation{Institute of Physics  ASCR, Cukrovarnick\'a 10, 162 53
Praha 6, Czech Republic }
\affiliation{School of Physics and Astronomy, University of Nottingham,
Nottingham NG7 2RD, UK}
\author{V. K. Dugaev}
\affiliation{Departamento de Fisica and CFIF, Instituto Superior Tecnico, Av. Rovisco Pais, 1049-001 Lisboa, Portugal}
\author{Jairo Sinova}
\affiliation{Department of Physics, Texas A\&M University,
College Station, TX 77843-4242, USA}

\date{\today}

\begin{abstract}
The anomalous Hall effect (AHE) is a consequence of spin-orbit coupling in a ferromagnetic metal and
related primarily to density-matrix response to an electric field that is off-diagonal in band index.
For this reason disorder contributions to the AHE
are difficult to treat systematically using a semi-classical Boltzmann equation
approach, even when weak localization corrections are disregarded. 
In this article we explicitly demonstrate the equivalence of an appropriately modified 
semiclassical transport theory which includes anomalous velocity and side jump contributions 
and microscopic Kubo-Streda perturbation theory, with particular unconventional contributions
in the semiclassical theory identified with particular Feynman diagrams when calculations are 
carried out in a band-eigenstate representation. The equivalence we establish is 
verified by explicit calculations for the case of the two-dimensional (2D) Dirac model Hamiltonian
relevant to graphene.  
\end{abstract}


\maketitle

\section{Introduction}

In spite of its long history as a basic ferromagnetic metal 
characterization tool, the theory of the anomalous Hall effect (AHE) 
continues to be a subject of confusion and debate.\cite{Sinova:2004_c} 
Theoretical descriptions of  the dc AHE invariably involve 
long and complex calculations and usually do not lead to 
simple general results with transparent interpretations.\cite{{Karplus:1954_a},{Kohn:1957_a},{Luttinger:1958_a},Sinova:2004_c} 
Theories of the AHE normally focus on particular simple model Hamiltonians,
and ignore interactions apart from mean-field exchange potentials that encode magnetic order.
Even with these simplifications,  the AHE {\em problem} tends to be 
difficult because the effect usually follows mainly from density-matrix linear 
response that is off-diagonal in Bloch state band indices, {\em i.e.} from 
induced interband coherence rather than simply from changes in Bloch state occupations. 
Farraginous theoretical results have followed from the application of different methods to 
the same models.\cite{{Dugaev:2005_a},{Sinitsyn:2005_a},{Liu:2005_c},{Inoue:2006_a},{Onoda:2006_a}}  
Rigorous approaches based on Green's function techniques, like the Keldysh
or Kubo-Streda formalisms, have the advantage of 
being systematic but can be technically more difficult to implement 
and are often not physically transparent.

Renewed interest in reaching consensus on a general theory of the AHE has been fueled in part by the
realization of diluted magnetic semiconductor (DMS) ferromagnets which have 
strong spin-orbit interactions, large anomalous Hall effects, and variable carrier 
densities and magnetic properties.  DMS ferromagnets can 
serve as a testing ground for the different theoretical predictions concerning the 
anomalous Hall effect.\cite{Jungwirth:2002_a,Dietl:2003_c,Jungwirth:2003_b,Chun:2006_a} 
Interest has also been increased by the (phenomenological) 
demonstration that a relatively simple {\em intrinsic} contribution that is a momentum-space Berry phase 
band-structure property dominates the AHE in many ferromagnets \cite{Taguchi:2001_a,Shindou:2001_a,Onoda:2002_a,Fang:2003_a,Yao:2004_a,Lee:2004_a,Kotzler:2005_a,Zeng:2006_a,Sales:2006_a}.

From one point of view, the main reason for the physical opaqueness of formal microscopic 
approaches to the AHE is that the Hall current is much weaker than the current parallel to the electric field.
No Hall contribution appears in usual theories of the dc conductivity which use Gaussian  
disorder potential distribution models and evaluate
conductivity to the leading order of the small parameter $\xi = 1/(k_Fl_{sc}),$
where $l_{sc}$ is the disorder scattering length.  Higher order terms must
therefore be considered - and these are unfortunately plentiful.  Bookkeeping becomes a challenge. 
Additionally typical metallic or semiconducting ferromagnets have many occupied bands and 
anisotropic band structures, making it difficult in general to obtain analytical solutions to the 
relevant transport equations.  

The semiclassical description of transport 
coefficients through a Boltzmann equation does lead to a Hall contribution if 
skew scattering\cite{{Loss},{Mott},{Smit:1955_a}} is accounted for in the collision term.
There are however, other contributions to the Hall effect that are ignored in such a theory
including anomalous velocity \cite{{Jungwirth:2002_a},{Sundaram:1999_a},{Fang:2003_a},{Yao:2004_a},{Blount-review}} 
and the side jump \cite{{Sinitsyn:2005_a},Adams:1959_a,{Berger:1970_a},{Nozieres:1973_a},{Chazalviel:1975_a},{Crepieux:2001_a},{Sinitsyn:2005_b}}
effects.  The presence of these additional contributions is directly linked to induced nonzero interband elements of the density matrix, in other words,
the interband coherence, 
as recognized since the work of Kohn and Luttinger. \cite{{Luttinger:1955_a},{Luttinger:1958_a}}

\begin{widetext}

%
\begin{figure}[h]
\includegraphics[width=16 cm]{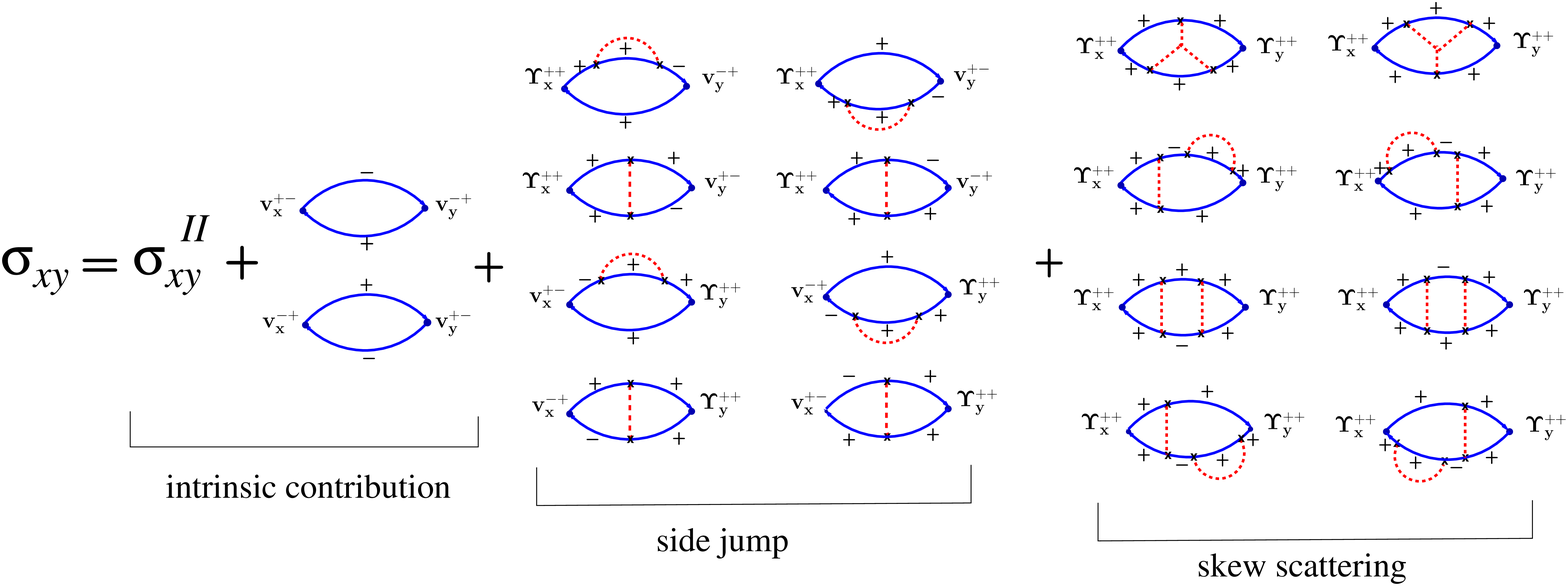}
\centering
\caption{Graphical representation of the AHE conductivity within the chiral 
(band eigenstate) representation.
The two bands of the two-dimensional Dirac model are labeled "$\pm$".
 The subsets of diagrams
that correspond to specific terms in the semiclassical Boltzmann formalism
are indicated. A detailed discussion of each set of diagrams is given in Sec. \ref{comp}.}
\label{mainfig}
\end{figure}

\end{widetext}

In a previous paper several of us demonstrated \cite{Sinitsyn:2005_b} that by utilizing the Fermi golden rule and a newly derived  
gauge invariant expression for the coordinate shift due to a scattering
event one can construct a semiclassical, physically transparent description of all 
contributions to the anomalous Hall effect and obtain 
the same final results as derived in a more complicated  way using the microscopic perturbative
approach of Luttinger.\cite{Luttinger:1958_a} The major drawback of this semiclassical approach is the 
fact that the identification of contributions to the AHE is not systematic; it is not obvious for 
example that other contributions of the same size are not being neglected and therefore, the assertion that 
we do include all main contributions must be verified by comparison to other approaches since the semiclassical 
approach lacks the rigor of the microscopic approaches. 
The main goal of the present paper is to provide a more detailed explanation of the semiclassical approach
and to link it with a more systematic and formally rigorous approach based on the Kubo-Streda formula. 
The various different contributions that have been identified in the semiclassical approach,
can, for the particular case of two-dimensional Dirac bands, all be directly linked to particular
Feynman-diagram subsums in the Kubo-Streda formalism.
The correspondence is summarized in Fig. \ref{mainfig}, in which the diagrammatic expansion
of the Kubo-Streda formalism in the chiral (band-eigenstate) basis is summarized.  
In Fig. \ref{mainfig} Feynman diagrams are grouped in the three major contributions that arise 
naturally in the semiclassical approach: intrinsic, side-jump, and skew-scattering. 

We organize the rest of the article as follows. In Sec.~\ref{MH} we introduce the model Hamiltonian for
which we demonstrate the equivalence of the two approaches.
In Sec. \ref{SBE} we provide a formal description of both techniques and
discuss the approximations involved in their application. 
In Sec. \ref{comp} we discuss different contributions identified in the Boltzmann approach:
intrinsic, side-jump, and skew scattering contributions, and show their equivalence to 
partial sums of diagrams in 
the Kubo-Streda formalism. In Sec. \ref{missing} we discuss some 
diagrams that appear in the Kubo expansion beyond our approximations.
Finally we summarize our results in Sec.~\ref{conclusion}.

\section{Model Hamiltonian}

\label{MH}
Although most of the discussion in the following sections is general and applicable to electronic structure
in any number of dimensions, we choose as a concrete example the massive 2D Dirac model 
Hamiltonian with randomly placed weak $\delta$-function-like spin-independent impurities.
We have chosen this model due to its simplicity
which permits the derivation of compact analytical expressions that 
facilitate comparison between the semiclassical Boltzmann approach and the microscopic Kubo approach.  
The impurity free 2D Dirac Hamiltonian is
\begin{equation}
\hat{H}_0=v(k_x \sigma_x +k_y \sigma_y) +\Delta \sigma_z.
\label{dh1}
\end{equation}
where $\sigma_x$ and $\sigma_y$ are Pauli matrices. 
(In the case of graphene, this degree-of-freedom represents the two sites per cell on a honeycomb lattice.)
Here and throughout the 
text we take $\hbar=1$. 
This Hamiltonian breaks time-reversal symmetry and therefore has a non-zero Hall conductivity.
Kubo formula calculations of the dc-limit of the Hall conductivity for this Hamiltonian have been already
performed\cite{Sinitsyn:2006_a} in the self-consistent non-crossing approximation
with the following result which can be applied to the 
charge and spin Hall conductivities in metallic graphene: 
\begin{eqnarray}
\sigma_{xy}&=-\frac{e^2\Delta }{4\pi \sqrt{(vk_F)^2+\Delta^2}} [1+
\frac{4(vk_F)^2}{4\Delta^2+(vk_F)^2}\nonumber\\&+
\frac{3(vk_F)^4}{(4\Delta^2+(vk_F)^2)^2} ]  -
  \frac{ e^2V_1^3 }{2\pi n_i V_0^4 } \frac{\ (vk_F)^4 \Delta}
{  (4\Delta^2+(vk_F)^2)^2},
\label{sxy2}
\end{eqnarray}
where $k_F$ is the Fermi momentum and the parameters $V_0$ and $V_1$ characterizing the disorder distribution as defined below.

\begin{figure}[h]
\includegraphics[width=5cm]{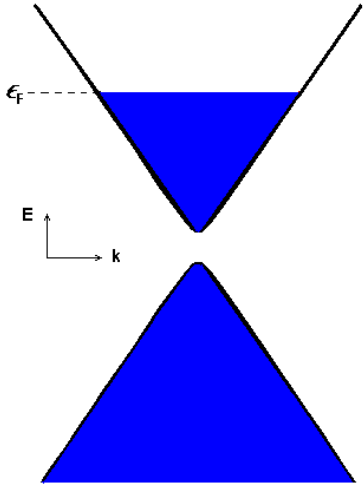}
\centering
\caption{Energy dispersion for the massive 2D Dirac Hamiltonian. Filled eigenstates are represented by the colored region. 
The Fermi level is assumed to be above the gap.}
\label{cone1}
\end{figure}

The non-zero mass $\Delta$ opens a gap in the spectrum and splits the Dirac band into sub-bands
above and below the gap with dispersions
\begin{equation}
\epsilon_{{\bf k}}^{\pm}=\pm \sqrt{\Delta^2+(v k)^2}
\label{ep1}
\end{equation}
where $k=|{\bf k}|$  and the labels $\pm$ distinguish bands with positive and negative energies.
In what follows, we will assume that the Fermi energy is positive and the Fermi level lies above the gap, as 
illustrated in Fig.~\ref{cone1}. 
%

The chiral basis states that diagonalize the Hamiltonian in Eq. (\ref{dh1}) are 
$\psi^{\pm}_{{\bf k}} =(1/\sqrt{V})e^{i{\bf k\cdot r}} |u^{\pm}_{{\bf k}}\rangle$, where 
\begin{equation} 
|u^{+}_{{\bf k}}\rangle=\left( \begin{array}{l}
\cos(\theta/2) \\
\sin(\theta /2) e^{i\phi} 
\end{array} \right),\,\,\,\, 
|u^{-}_{{\bf k}}\rangle=\left( \begin{array}{l}
\sin(\theta/2) \\
-\cos(\theta /2) e^{i\phi} 
\end{array} \right),
\label{basis}
\end{equation}
$\cos(\theta)=\Delta/\sqrt{(vk)^2+\Delta^2}$,
and $\tan(\phi) = k_y/k_x$.
In the eigenstate representation (Eq.(\ref{basis})), the disorder and velocity operators have the
following matrix form
\begin{equation}
\hat{v}_x=v \left( \begin{array}{cc}
\sin(\theta)\cos(\phi) & -i \sin(\phi)-\cos(\theta)\cos(\phi) \\
i \sin(\phi) -\cos(\theta)\cos(\phi) & -\sin(\theta) \cos(\phi)
\end{array} \right),
\label{sx1}
\end{equation}
\begin{equation}
\hat{v}_y=v \left( \begin{array}{cc}
\sin(\theta)\sin(\phi) & i \cos(\phi)-\cos(\theta) \sin(\phi) \\
-i \cos(\phi) -\cos(\theta) \sin(\phi) & -\sin(\theta) \sin(\phi)
\end{array} \right), 
\label{sy1}
\end{equation}
and 
\begin{equation}
\begin{array}{l}
V_{{\bf k'k}}=V^0_{{\bf k'k}} \left( \begin{array}{ll}
\langle u_{{\bf k'}}^{+} |u_{{\bf k}}^{+} \rangle & \langle u_{{\bf k'}}^{+} |u_{{\bf k}}^{-} \rangle \\
\langle u_{{\bf k'}}^{-} |u_{{\bf k}}^{+} \rangle & \langle u_{{\bf k'}}^{-} |u_{{\bf k}}^{-} \rangle
\end{array} \right)=V^0_{{\bf k'k}}\times \\
\\
\times\left( \begin{array}{ll}
1-\sin^2(\frac{\theta}{2}) (1-e^{i(\phi-\phi')}) & \sin(\theta)  \frac{1-e^{i(\phi-\phi')}}{2} \\
\sin(\theta) \frac{1-e^{i(\phi-\phi')}}{2} & 1-\cos^2(\frac{\theta}{2})(1- e^{i(\phi-\phi')})
\end{array} \right)
\end{array}
\label{vkk1}
\end{equation}
where $V^0_{{\bf k'k}}$ orbital disorder matrix element.  

We consider the model of randomly located $\delta$-function
scatterers: $V({\bf r}) =\sum_i V_i \delta ({\bf r}-{\bf R}_i)$ with 
$R_i$ random and strength distributions satisfying $\langle V_i \rangle_{dis} =0$, $\langle V_i^2 \rangle_{dis} =V_0^2 \ne 0$ 
and $\langle V_i^3 \rangle_{dis} = V_1^3 \ne 0$. 
This model is different from the standard 
white noise disorder model in which 
only the second order cumulant is nonzero; 
$\langle |V^0_{{\bf k'k}}|^2 \rangle_{dis} = n_i V_0^2 $ where $n_i$ is the impurity concentration and other
correlators are either zero or related to this correlator by Wick's theorem.
The deviation from white noise in our disorder model is quantified by $V_1\ne 0$,
and is necessary to capture part of the skew scattering contribution
to the anomalous Hall effect. Note however that within this model, as we will show in later sections, there are other 
contributions to the skew scattering that arise from the Gaussian disorder distribution which turn out to be independent of disorder strength,
e.g. see Sec.~\ref{skew_section}. 

In what follows we use $+$/$-$ for 
the upper/lower chiral band indices.  For example, from Eq. (\ref{sx1}) $v_x^{+-}= v[-i \sin(\phi)-\cos(\theta)\cos(\phi)]$ 
or from Eq. (\ref{vkk1}) 
$\langle |V_{{\bf k'k}}^{++}|^2 \rangle_{dis}=n_iV_0^2 |1-\sin^2(\frac{\theta}{2}) (1-e^{i(\phi-\phi')}) |^2$. 
Working in the eigenstate basis, we will refer to the velocity matrix elements $v^{++}_{x/y}$ and $v^{--}_{x/y}$ as 
the diagonal velocities and to  
$v^{+-}_{x/y}$ and $v^{-+}_{x/y}$ as the off diagonal or interband velocities.

\section{The Boltzmann equation and the Kubo-Streda formula}
\label{SBE}
\subsection{The semiclassical Boltzmann equation.}

There is a substantial literature\cite{{Smit:1955_a},{Jungwirth:2002_a},{Sinitsyn:2005_a},{Sinitsyn:2005_b}}
on the application of Boltzmann equation concepts to AHE theory. 
However, stress was usually 
placed only on one of the many possible mechanisms for a Hall current.
In this section we briefly review this approach, 
incorporating some new insights from recent work.\cite{Sinitsyn:2005_b}

The semi-classical Boltzmann equation (SBE) describes the evolution of the electron
distribution function as though electrons were classical particles labeled by a 
band index and living in the crystal's momentum space. From 
a quantum mechanical point of view, it is clear that this approach 
is not universally applicable because it violates the Heisenberg 
uncertainty principle by having a momentum-distribution
function at each point in space. A rigorous treatment of the electronic state evaluation in the presence of 
disorder should generally consider the entire density matrix, including off-diagonal elements. 
Formal justifications of the Boltzmann description are usually made in terms of the properties of 
wave packets that have well defined average momentum, center of mass coordinate, spin, intrinsic angular momentum, {\em etc}.\cite{culcer}
The distribution functions then acquire the meaning of the probability distribution of wave packets
that behave in many respects like classical particles.\cite{Marder:1999_a}

The semi-classical distribution function evolves both due to hydrodynamic particle fluxes
and due to collisions with impurities. 
Quantum mechanically, the elementary scattering process is described by the scattering matrix from one state to another in the presence of 
a perturbing potential. 
In the semi-classical approach, scattering can be accompanied by changes of coordinate.
As we discuss below, these have to be taken into account when constructing a 
kinetic equation for the semi-classical distribution function that captures the AHE. 

\subsubsection{The golden rule.}

The golden rule connects the classical and quantum descriptions of a scattering event.\cite{Scattbook} It shows how the
scattering rate  $\omega_{l'l}$ between states with different quantum numbers $l$ and $l'$
is related to the so called $T$-matrix elements:
\begin{equation}
\omega_{l'l}=2\pi |T_{l'l}|^2 \delta(\epsilon_{l'}-\epsilon_{l}).
\label{WT}
\end{equation}
The scattering $T$-matrix is defined as
\begin{equation}
T_{l'l}=\langle l'| \hat{V}| \psi_{l} \rangle
\label{Tmatrix}
\end{equation}
where $\hat{V}$ is the impurity potential operator and 
$| \psi_{l} \rangle$ is the eigenstate of the full Hamiltonian 
$\hat{H}=\hat{H}_0+\hat{V}$ that satisfies the Lippman-Schwinger 
\begin{equation}
| \psi_{l} \rangle = |l\rangle  +\frac{\hat{V}}{\epsilon_{l}-\hat{H}_0+i\eta} | \psi_{l} \rangle. 
\label{psil}
\end{equation}
For weak disorder one can approximate the scattering state $| \psi_{l} \rangle$ 
by a truncated series in powers of $V_{ll'}=\langle l|\hat{V} |l'\rangle$ and
take $l=({\bf k},\mu)$ as the combined (momentum,band) index of the eigenstate $|l\rangle=|{\bf k},\mu\rangle$ of
the unperturbed  Hamiltonian $\hat{H}_0$: 
\begin{equation}
| \psi_{l} \rangle \approx |l\rangle +\sum_{l''} \frac{V_{l''l}} {\epsilon_{l}-\epsilon_{l''}+i\eta} | l''\rangle + \ldots
\label{ser1}
\end{equation}
For example the $T$-matrix up to the second order in $\hat{V}$ is
\begin{equation}
T_{l'l}\approx V_{l'l}+\sum_{l''} \frac{V_{l'l''}V_{l''l}}{\epsilon_{l}-\epsilon_{l''}+i\eta}.
\label{Tmatrix2}
\end{equation}
Substituting Eq. (\ref{Tmatrix2}) into Eq. (\ref{WT}) one can find the scattering rate up to the 3rd order in the disorder strength
 \begin{equation}
 \omega_{ll'}=\omega^{(2)}_{ll'}+\omega^{(3)}_ {ll'}+\cdots,
 \label{om1}
 \end{equation}
where
\begin{equation}
\omega^{(2)}_{ll'}=2\pi \langle |V_{ll'}|^2 \rangle_{dis} \delta (\epsilon_{l} -\epsilon_{l'})
\label{om2}
\end{equation}
\begin{equation}
\omega^{(3)}_ {ll'}=2\pi \left ( \sum_{l''} \frac{\langle V_{ll'}
 V_{l'l''} V_{l''l}\rangle_{dis}}
{\epsilon_{l} -\epsilon_{l''}-i \eta} +c.c. \right) \delta (\epsilon_{l} -\epsilon_{l'}).
\label{om3}
\end{equation}

The skew scattering contribution to the Hall effect follows from the asymmetric part of the scattering rate:\cite{Smit:1955_a}
\begin{equation}
\omega_{ll'}^{(a)} \equiv \frac{\omega_{ll'} - \omega_{l'l}}{2}.
\label{ss1}
\end{equation}
Since $\omega^{(2)}_{ll'}$ is symmetric, the leading contribution to $\omega_{ll'}^{(a)}$ 
appears at order $V^{3}$.
The $\omega^{(3)}_{ll'}$ contribution to the scattering rate $\omega_{ll'}$ contains symmetric and antisymmetric parts. 
The symmetric part is not essential since it only
renormalizes the second order result of Eq.~(\ref{om2}). 
The sum  in parentheses in Eq.~(\ref{om3}) can be rewritten in the form
\begin{equation}
\begin{array}{l}
P \left( \sum_{l''}  \frac{2{\rm Re}\langle V_{ll'} V_{l'l''} V_{l''l}\rangle_{dis}}
{\epsilon_{l} -\epsilon_{l''}} \right)  - \\
\\
2\pi \sum_{l''} \delta (\epsilon_{l}-\epsilon_{l''})
{\rm Im} \langle V_{ll'} V_{l'l''} V_{l''l}\rangle_{dis}
\end{array}
\label{sum3}
\end{equation}
The first term is symmetric under the exchange of indexes  $l \leftrightarrow l'$
(note that this needs the fact that $\epsilon_{l} = \epsilon_{l'}$ due to the delta-function in Eq.~(\ref{om3})), hence this sum does not contribute to the asymmetric
part of the scattering amplitude and one can concentrate on the second one:
\begin{equation}
\begin{array}{l}
\omega^{(3a)}_ {ll'}= -(2\pi)^2  \sum \limits_{l''} \delta (\epsilon_{l}
 -\epsilon_{l''}) {\rm Im} \langle V_{ll'} V_{l'l''} V_{l''l}\rangle_{dis} 
 \delta (\epsilon_{l} -\epsilon_{l'}),
\end{array}
\label{omasym1}
\end{equation}
with the superscript $3a$ meaning that this is the antisymmetric part of the scattering rate calculated 
at order $V^3$.

\subsubsection{The coordinate shift}
In a semi-classical description of wavepacket motion in a crystal, scattering produces both a change 
in the direction of motion and a separate coordinate shift.\cite{{Adams:1959_a},{Berger:1970_a},{Nozieres:1973_a},{Sinitsyn:2005_b},{Sinitsyn:2005_a}}
In the lowest order Born approximation for a scalar disorder potential one can derive an expression for the
coordinate shift which accompanies scattering between two band states. The shift
does not depend explicitly on the type of impurity and can be expressed in terms of  
initial and final states only:\cite{{Sinitsyn:2005_a},{Sinitsyn:2005_b}}
\begin{equation}
 \delta {\bf r}_{l'l} = \langle u_{l'}| i\frac{\partial}{\partial {\bf k'}} u_{l'} \rangle - 
 \langle u_{l}| i\frac{\partial}{\partial {\bf k}} u_{l} \rangle
 - \hat{{\bf D}}_{{\bf k',k}} {\rm arg}[\langle u_{l'}|u_{l}\rangle].
\label{delr4b}
\end{equation}
where ${\rm arg}[a]$ is the phase of the complex number $a$ and
$$
\hat{{\bf D}}_{{\bf k',k}}= \frac{\partial}{\partial {\bf k'}} + \frac{\partial}{\partial {\bf k}}
$$
The topological interpretation of Eq. (\ref{delr4b}) has been explained in Ref.~\onlinecite{Sinitsyn:2005_b}. The first two terms in Eq. 
(\ref{delr4b}) have been known for a long time.  The last term was first derived only recently in Ref.~\onlinecite{Sinitsyn:2005_b} 
but is an essential contribution which makes the expression for the coordinate shift gauge invariant.

\subsubsection{Kinetic equation of the semi-classical Boltzmann distribution}
Eqs. (\ref{WT}) and (\ref{delr4b}) contain the quantum mechanical information necessary to write down a semi-classical Boltzmann
equation that takes into account both the change of momentum and the coordinate shift during scattering in a homogeneous crystal 
in the presence of a driving electric field $\bf E$. Keeping only terms up to the linear order in the electric field the Boltzmann
equation reads\cite{Sinitsyn:2005_b}
\begin{equation}
\begin{array}{l}
  \frac{\partial f_l}{\partial t} +e{\bf E}\cdot{\bf v}_{0l} \frac{\partial f_{0} (\epsilon_l) }{\partial {\epsilon_l}}
  = -\sum_{l'}
   \omega_{ll'} [ f_{l}-\\
\\
-f_{l'}-\frac{\partial f_{0} (\epsilon_l) }{\partial {\epsilon_l}}e{\bf E} \cdot \delta {\bf r}_{l'l}],
\end{array} 
\label{beint}
 \end{equation}
 where ${\bf v}_{0l}$ is the usual velocity 
\begin{equation}
{\bf v}_{0l} = \partial \epsilon_l/\partial {\bf k}.
\label{vol12}
\end{equation}
Note that if only elastic scatterings with static impurities are responsible for the collision term,
the rhs. of Eq. (\ref{beint}) is linear in $f_{l}$ and not in $f_{l} (1-f_{l'})$. 
This property follows from the fact that Heizenberg
time evolution equations for 
creation or annihilation operators in a noninteracting electron system are
linear and can be mapped to the Schr\"odinger equation for amplitudes in a single
particle system. (For further discussion of this point see Appendix B in Ref. \onlinecite{Luttinger:1955_a}.)

This Boltzmann equation has the standard form except for the coordinate shift effect which 
is taken into account in the last term in the collision integral on the rhs of Eq. (\ref{beint}).
This term appears because the kinetic energy of a particle in the state
$l'$ before scattering into the state $l$ is smaller than $\epsilon_l$ by the amount
$e{\bf E} \cdot \delta {\bf r}_{ll'}=-e{\bf E} \cdot \delta {\bf r}_{l'l}$.   The collision term
does not vanish in the presence of an electric field when the occupation probabilities $f_l$ are 
replaced by their thermal equilibrium values because 
$f_{0}(\epsilon_l)-f_{0}(\epsilon_l-e{\bf E} \cdot \delta {\bf r}_{ll'}) \approx -\frac{\partial f_{0} 
(\epsilon_l) }{\partial {\epsilon_l}}e{\bf E} \cdot \delta {\bf r}_{l'l} \ne 0$. 
The last term in the collision integral in Eq. (\ref{beint})\cite{Sinitsyn:2005_a} accounts for 
this interplay between coordinate shifts and spatial variation of local chemical potential 
in the presence of an electric field. 
 
The total distribution function $f_{l}$ in the steady state ($\partial f_l/\partial t =0$) 
can be written as the sum of the equilibrium distribution $f_{0}(\epsilon_l)$ and non-equilibrium contributions, 
\begin{equation}
f_{l}=f_{0}(\epsilon_l)+g_{l}+g_l^{adist}
\label{ffgg}
\end{equation}
where we have split the non-equilibrium contribution into two terms $g_l$ and $g^{adist}$
which solve independent self-consistent time-independent equations:\cite{Sinitsyn:2005_b}
\begin{equation}
 e{\bf E}\cdot{\bf v}_{0l} \frac{\partial f_{0} (\epsilon_l) }{\partial {\epsilon_l}}
  = -\sum_{l'} \omega_{ll'} (g_{l}-g_{l'} )
\label{bbeint3}
\end{equation}
and
\begin{equation}
\sum_{l'}\omega_{ll'}  \left( g^{adist}_l - g^{adist}_{l'} +\frac{-\partial f_0(\epsilon_l)}{\partial \epsilon_l} e{\bf E} \cdot  \delta {\bf r}_{l'l} \right) =0.
\label{ganl}
\end{equation}
(The label {\em adist} stands for anomalous distribution.) 
 A standard approach to solving these equations in 2D is to look for the solution of
 Eq. (\ref{bbeint3}) in the form (see e.g. Ref. \onlinecite{Loss}):
 \begin{equation}
 g_{l}=(-\frac{df_{0} (\epsilon_l)}{d\epsilon_l})e{\bf E}\cdot \left( A_{\mu} {\bf v}_{0l} +B_{\mu} {\bf v}_{0l} \times {\bf \hat{z}} \right),
 \label{solg}
 \end{equation}
 where ${\bf \hat{z}}$ is the unit vector in the out-of-plane direction.
 We will assume that the transverse conductivity is much smaller than the longitudinal one so that $A_{\mu}>>B_{\mu}$. 
 One then finds by  the direct substitution of Eq. (\ref{solg}) into Eq. (\ref{beint}) that
\begin{equation}
A_{\mu}=\tau_{\mu}^{tr},\,\,\,\,\,\, B_{\mu}=(\tau_{\mu}^{tr})^2/\tau_{\mu}^{\perp}
\label{ABtaus}
\end{equation}
where
 \begin{equation}
 \begin{array}{l}
 1/\tau_{\mu}^{tr}=\sum_{\mu'} \int \frac{d^2{\bf k^{\prime}}}{(2\pi)^2}
  \omega_{ll'} (1-\frac{|v_{l'}|}{|v_{l}|} \cos (\phi-\phi'))
 \\
 \\
 1/\tau_{\mu}^{\perp}=\sum_{\mu'} \int \frac{d^2{\bf k^{\prime}}}{(2\pi)^2}  \omega_{ll'}\frac{|v_{l'}|}{|v_{l}|} \sin (\phi-\phi')
 \end{array}
 \label{taus}
 \end{equation}
and $\phi$ and $\phi'$ are the angles between ${\bf v}_{0l}$ or ${\bf v}_{0l'}$ and the x-axes. 
This completes the solution of Eq. (\ref{bbeint3}) for $g_l$. 
We will show in Sec.III.B.1 how a similar {\em ansatz} can also solve Eq. (\ref{ganl}) for the 
anomalous distribution $g_l^{adist}$.

\subsubsection{Modified velocities and the anomalous Hall effect.}

To find the current induced by an electric field and hence the conductivity
we need to derive an appropriate expression for the velocity of semiclassical particles,
in addition to solving the SBE for the state occupation probabilities.   In considering the 
AHE, in addition to the band state group velocity ${\bf v}_{0l}=\partial \epsilon_l /\partial {\bf k}$ 
one should also take into account velocity renormalizations due to the accumulations of 
coordinate shifts after many scattering events (the side-jump effect) and due to 
band mixing by the electric field (the anomalous velocity effect):\cite{{Nozieres:1973_a},{Sinitsyn:2005_b}}
\begin{equation}
{\bf v}_l=\frac{\partial \epsilon_l }{\partial {\bf k}} + {\bf F}_l  \times e {\bf E} + \sum_{l'} \omega_{l'l} \delta {\bf r}_{l'l}.
\label{tovel}
\end{equation}
The second term in Eq. (\ref{tovel})  captures changes in the speed at which
a wave packet moves between  scattering events under the influence of the  external electric field \cite{Appendix} and
${\bf F}_l$ is the Berry curvature of the band \cite{Sundaram:1999_a}
\begin{equation}
 ( F_l)_k = \epsilon_{ijk} {\rm Im} \left[ \langle \frac{\partial u_{l}} {\partial  k_j}| \frac{\partial u_{l}}{\partial k_i}\rangle\right].
\label{berrcurv}
\end{equation}
The last term in Eq. (\ref{tovel}) is due to the accumulation of coordinate shifts 
after many scatterings. Combining Eqs. (\ref{ffgg}) and (\ref{tovel}) we obtain the total current 
\begin{equation}
{\bf j} = e \sum_l f_l {\bf v}_l
\label{jtotal}
\end{equation}

\subsubsection{Mechanisms of the AHE.} 

From (\ref{jtotal}) we can extract the off-diagonal Hall conductivity which is naturally written 
as the sum of four contributions:
\begin{equation}
\sigma_{xy}^{total}=\sigma_{xy}^{int}+\sigma_{xy}^{adist}+\sigma_{xy}^{sj}+\sigma_{xy}^{sk}.
\label{jttt}
\end{equation}
The first term is the so called intrinsic contribution
\begin{equation}
\sigma_{xy}^{int} = -e^2 \sum_l f_0(\epsilon_l) F_l,
\label{jtotal_i}
\end{equation}
which is due to the anomalous velocity of free electrons under the action of the electric field
 \begin{equation}
 {\bf v}^{(a)}_l={\bf F}_l  \times e {\bf E}.
 \label{anvel}
 \end{equation}
This term
is called the intrinsic contribution because it is not related to impurity scattering, {\em i.e.} 
its evaluation does not require knowledge of the disorder present in the system.  The intrinsic AHE  
is completely determined by the topology of the Bloch band. 

The next two contributions follow from coordinate shifts during scattering events:
\begin{equation}
\sigma_{xy}^{adist} = e \sum_l( g_l^{adist}/E_y) (v_{0l})_x
\label{jtotal_adist}
\end{equation}
is the conductivity
 due to the anomalous correction to the distribution function 
while the direct side-jump contribution 
\begin{equation}
\sigma_{xy}^{sj} =  e\sum_l (g_l/E_y) \sum_{l'} \omega_{l'l} (\delta {\bf r}_{l'l})_x
\label{jtotal_sj}
\end{equation}
is due to the side-jump velocity, {\em i.e.} due to the accumulation of coordinate shifts after many scattering
events. Since coordinate shifts are responsible both for $\sigma_{xy}^{adist}$ and for $\sigma_{xy}^{sj}$ 
there is, unsurprisingly, an intimate relationship between those two contributions.
We will demonstrate their equality by evaluating $\sigma_{xy}^{adist}$ below. 
Because of this relationship, in most of the literature, $\sigma_{xy}^{adist}$ is 
usually considered to be part of the side-jump contribution, i.e. $\sigma_{xy}^{adist}+\sigma_{xy}^{sj}\rightarrow \sigma_{xy}^{sj}$.
Here we will distinguish between the two when connecting the SBE approach to the Kubo-Streda formalism. 

Finally, $\sigma_{xy}^{sk}$ is the skew scattering contribution
\begin{equation}
\sigma_{xy}^{sk} =  e\sum_l (g_l/E_y) (v_{0l})_x.
\label{jtotal_sk}
\end{equation}
The skew scattering contribution is independent of the coordinate shift and of the anomalous velocity. It is non-zero 
for an isotropic energy dispersion $\epsilon_l$ only when the scattering rate $\omega_{ll'}$ is asymmetric. 
Writing the transverse velocity in the form $(v_{0l})_x=|v_{0l}|\cos(\phi)$,
then using Eq.~(\ref{taus}) and substituting Eq. (\ref{solg}) for $g_l$ into Eq. (\ref{jtotal_sk})
one finds
 \begin{equation}
 \begin{array}{l}
 \sigma_{xy}^{sk}=e\sum_{\mu} \int \frac{d^2{\bf k}}{(2\pi)^2}  (g_{l}/E_y) |v_{0l}| \cos(\phi) = \\
 \\
 =-e^2\sum_{\mu} \frac{(\tau_{\mu}^{tr})^2}{\tau_{\mu}^{\perp}} \int \frac{d^2{\bf k}}{(2\pi)^2}
   (-\frac{df_{0}(\epsilon_l)}{d\epsilon_l}) |v_{l}|^2 \cos ^2(\phi).
 \end{array}
 \label{cond1}
 \end{equation}
 At zero temperature this expression simplifies to
 \begin{equation}
 \begin{array}{l}
 \sigma_{xy}^{sk}=-\sum_{\mu} \frac{ e^2 (\tau_{\mu}^{tr})^2  v_{F\mu} k_{F\mu} }   {4\pi \tau_{\mu}^{\perp}}
 \end{array}
 \label{cond11}
 \end{equation}

There is potentially one more
contribution which has not been included in Eq.~(\ref{jttt}) and which involves the product of $g^{adist}$ and 
the side jump velocity $\sum_{l'}\omega_{l'l}\delta{\bf r}_{l'l}$. Such a term
 does not contribute 
to the Hall conductivity in the present model calculation of isotropic scattering and band structure. It may however, in principle,
give rise to a non-zero contribution given an appropriate anisotropic scattering and anisotropic Fermi distribution. We do not 
consider this term further here.

In Sec. IV we evaluate each of these four contributions for the particular example of the 2D Dirac model.

\subsection{The Kubo-Streda formula.}

A fully quantum mechanical formally exact expression for the
conductivity 
is provided by the linear response Kubo theory.
There are various equivalent formulations of this approach. In one version
the conductivity at zero temperature $T$ is expressed in terms
of causal Green's functions.  
This formalism has previously been applied to the 
AHE in  an electron system with Rashba spin-orbit coupling.\cite{Dugaev:2005_a}
In the present paper we work mainly with another 
 version of the Kubo formula,
derived specially for the 
the dc conductivity. Its advantage is that one should not perform the 
$\omega \rightarrow 0$ limit at the end of calculations.   

The starting point in this approach is the Bastin formula:\cite{Crepieux:2001_a}
\begin{eqnarray}
\sigma_{xy}&=&-i\frac{e^2}{V}\int_{-\infty}^\infty \frac{d\epsilon}{2\pi i} f(\epsilon)
{\rm Tr}[-\hat{v}_x \delta(\epsilon-H) \hat{v}_j\frac{dG^R}{d\epsilon}\nonumber\\&&+
\hat{v}_i\frac{dG^A}{d\epsilon}\hat{v}_y\delta(\epsilon-H)],
\end{eqnarray}
which can be manipulated to give the Kubo-Streda formula for the dc $T=0$ Hall conductivity
$\sigma_{xy}=\sigma_{xy}^{I(a)}+\sigma_{xy}^{I(b)}+\sigma_{xy}^{II}$ where:
\begin{equation}
\sigma_{xy}^{I(a)}=\frac{e^2}{2\pi V} {\rm Tr} \langle \hat{v}_x G^R(\epsilon_F) \hat{v}_y G^A(\epsilon_F)\rangle_c,
\label{sigmaIa}
\end{equation}
\begin{equation}
\sigma_{xy}^{I(b)}=-\frac{e^2}{4\pi V} {\rm Tr} \langle \hat{v}_x G^R(\epsilon_F) \hat{v}_y G^R(\epsilon_F)+\hat{v}_i G^A(\epsilon_F) \hat{v}_j G^A(\epsilon_F)\rangle_c,
\label{sigmaIb}
\end{equation}
\begin{eqnarray}
\sigma_{xy}^{II}&=&
\frac{e^2}{4\pi V} \int_{-\infty}^{+\infty} d\epsilon f(\epsilon) {\rm Tr}[v_x G^R(\epsilon) v_y \frac{G^R(\epsilon)}{d\epsilon} -
\nonumber\\&&
-v_x \frac{G^R(\epsilon)}{d\epsilon}v_yG^R(\epsilon)-
v_x G^A(\epsilon) v_y \frac{G^A(\epsilon)}{d\epsilon} + \nonumber\\
&&+v_x\frac{G^A(\epsilon)}{d\epsilon}v_yG^A(\epsilon)]
=ec 
\left. \frac{\partial N(\epsilon)}{\partial B}
\right|_{\epsilon=\epsilon_F, B=0}
\label{sigmaII}
\end{eqnarray}
here the subscript $c$ indicates a disorder configuration average. 
The last contribution, $\sigma^{II}_{xy}$, was first derived by Streda. \cite{Streda} 
In these equations $B$ is an external magnetic field that couples only to the orbital degrees of freedom,
and $N(E)$ is the integrated density of states of the disordered system and 
$G^{R/A}(\epsilon_F)=(\epsilon_F-H\pm i\delta)^{-1}$ are the retarded and advanced Green's functions evaluated
at the Fermi energy. 
Every term in $\sigma_{xy}^{II}$ depends on products of retarded Green's functions only
or products of advanced Green's functions only. 
Since every such term has poles on the same side of the imaginary plane, disorder corrections to
this contribution are small because they contain at least one (small) factor $n_i V_0^2$
which is not 
compensated  by energy and momentum integrations.
Hence only the disorder free part of $\sigma_{xy}^{II}$ is important in the weak disorder limit, {\em i.e.} 
this contribution is zeroth order in the parameter $\xi=1/k_F l_{sc}$. It can also be shown by a similar argument that in general 
$\sigma_{xy}^{Ib}$, is of order $\xi$ and can be neglected in the weak scattering limit.\cite{Mahan}
Thus, important disorder effects are contained only in $\sigma_{xy}^{Ia}$.  In what follows we neglect 
$\sigma_{xy}^{Ib}$ and retain only the clean limit $\sigma_{xy}^{II}$ contribution.
The following paragraphs concern only $\sigma_{xy}^{Ia}$.

The effect of disorder on the disorder-configuration averaged Green's function is conveniently 
captured by the use of the T-matrix, defined by the integral equation $T=W+W G_0 T$ where $W=\sum_i V_0\delta(r-r_i)$. From this one obtains 
\begin{equation}
G=G_0+G_0 T G_0=G_0+ G_0 \Sigma G.
\label{G}
\end{equation}
  Upon disorder averaging we obtain 
\begin{equation}
\Sigma=\langle W\rangle_c+\langle W G_0 W\rangle_c
+\langle W G_0 W G_0 W\rangle_c +...
\end{equation}
To linear order in $n_i$ this translates to
\begin{equation}
\Sigma(z,\kk)= n_i V_{\kk,\kk}+\frac{n_i}{V}\sum_\kk V_{\kk,\kk'} G_0(\kk',z) V_{\kk',\kk},
\end{equation}
with $V_{\kk,\kk'}=V(\kk-\kk')$ being the Fourier transform of the single impurity
potential, which in the case of delta scatterers is simply $V_0$. 

Within the T-matrix formalism in the normal spin-basis we will take $V_1=0$ for simplicity since, as it will be shown when
considering the Kubo formalism within the chiral basis, only a couple of diagrams contribute to part of the skew scattering for $V_1\ne0$
and can therefore its contribution can be computed separately (see Sec.~\ref{skew_section}).

Replacing the bare Green's function by the self consistent Born approximation Green's function, Eq. (\ref{G}), in
the expression for $\sigma_{xy}^{Ia}$  
does not capture all the contributions to lowest order in $\xi$. One must also compute the ladder diagram correction
to the bare velocity vertex defined by
$\tilde{v}_\alpha(\epsilon_F)\equiv v_\alpha+\delta\tilde{v}_\alpha(\epsilon_F)$, where $\delta\tilde{v}_\alpha(\epsilon_F)$ satisfies
\begin{eqnarray}
\delta\tilde{v}_\alpha(\epsilon_F)=\frac{n_i V_0^2}{V}\sum_\kk G^R(\epsilon_F) v_\alpha G^A(\epsilon_F)\nonumber\\
+\frac{n_i V_0^2}{V}\sum_\kk G^R(\epsilon_F) \delta\tilde{v}_\alpha(\epsilon_F) G^A(\epsilon_F).
\end{eqnarray}
The explicit calculation of this vertex correction in the  $\sigma_z$ eigenstate basis
is described in Appendix \ref{vertex_calculation}.  The final result for the 2D Dirac model Hamiltonian is 
\begin{equation}
\tilde{v}_y=8v\Gamma \cos\theta\frac{(1+\cos^2\theta)}{\lambda(1+3\cos^2\theta)^2}\sigma_x
+\left(v+v\frac{\sin^2\theta}{(1+3\cos^2\theta)}\right)\sigma_y,
\end{equation}
where $\Gamma=\pi n_i^2V_0^2/4v^2$.
Incorporating this result in $\sigma_{xy}^{Ia}$ we obtain for
$\epsilon_F> \Delta$ 
\begin{equation}
\begin{array}{l}
\sigma_{xy}=\frac{e^2}{2\pi V}\sum_\kk {\rm Tr}\left[ v\sigma_x G \tilde{v}_yG \right] \\
\\
=-\frac{e^2 \cos\theta}{4\pi}\left(1+\frac{3\sin^2\theta}{(1+3\cos^2\theta)} + \frac{4\sin^2\theta}{(1+3\cos^2\theta)^2}\right)
\end{array}
\label{eq48}
\end{equation}
This is equivalent to the part of Eq.~(\ref{sxy2}) that is independent of $V_1$.
In order to understand more fully the relationship between this result and the 
more physically transparent SBE approach, we need to analyze the Kubo 
approach in the chiral band eigenstate basis. 
As described in the remaining part of this section, we are able to 
make the connection very clear.

We have also obtained the same result using the time ordered Green's functions formalism.  The
equivalence of the two results is shown in detail in Appendix~\ref{Dugaev}.

\subsubsection{Kubo-Streda approach in the band eigenstate basis.}

In the chiral Bloch eigenstate basis, specified in Eq. (\ref{basis}),
the disorder free Green's functions in momentum space can be written in the form
\begin{equation}
G^{R/A}_0(\epsilon)=
\frac{|u^{+}_{{\bf k}}\rangle \langle u^{+}_{{\bf k}}|}{\epsilon -\epsilon^{+}_{{\bf k}} \pm i\eta}+
\frac{|u^{-}_{{\bf k}}\rangle \langle u^{-}_{{\bf k}}|}{\epsilon -\epsilon^{-}_{{\bf k}}\pm i\eta}.
\label{gra1}
\end{equation}
From a calculational point of view the bare Green's function is simpler in this 
basis, but we must pay 
the price in the disorder matrix of the
 Hamiltonian which will contain, as seen in Sec. II,  intraband 
($ V^{++}_{{\bf kk'}}$, $V^{--}_{{\bf kk'}}$) and interband ($ V^{+-}_{{\bf kk'}}$, $V^{-+}_{{\bf kk'}}$) 
matrix elements. 

In the weak disorder limit, the energy width of Bloch state spectral peaks is smaller 
than the gap, allowing us to ignore direct interband scattering.
This does not mean, however, that interband matrix elements of the disorder potential are not important and can be disregarded.
In the rest of the text we will show that interband disorder matrix elements, and the interband 
coherence they imply, are in fact crucial to fully explain disorder contributions to the AHE.
Although interband scattering probability remains zero, interband disorder matrix elements
produce virtual transitions that mix states in the two bands in a way which is ultimately relevant. 

When evaluating diagrams in the chiral basis, disorder lines associated with interband and intraband scattering 
processes can be distinguished.
Their separate contributions will be easily linked to the semiclassical
interpretation of the scattering process.
As we will see below, the relevant vertex correction infinite subsum involves
only intraband elements $ V^{++}_{{\bf kk'}}$ and does not produce unusual effects in the 
sense that it only renormalizes the quantum life time and the diagonal velocity vertex, $v^{++}\rightarrow \Upsilon^{++}$. 
In contrast, vertex correction diagrams which contain $ V^{+-}_{{\bf kk'}}$, $V^{-+}_{{\bf kk'}}$ 
and therefore connect Green's functions with different band indices, do not have to be summed to infinite order and 
do capture the unusual transport physics which leads to AHE contributions.  
Only a few low order (in terms of $\xi$) diagrams,  which depend on $ V^{+-}_{{\bf kk'}}$ or $V^{-+}_{{\bf kk'}}$, have to be included in the calculations of $\sigma_{xy}$.
Note that these terms still contain the renormalized $\Upsilon^{++}$
 velocity vertex but the interband component does not contribute
to this renormalization as we will see below. 

\subsubsection{The self-energy in the non-crossing approximation.}

We consider first all possible diagrams that do not contain an 
interband part of the disorder potential.  This part of the calculation 
is essentially the familiar standard calculation because the 
Green's functions remain diagonal in band indexes and the calculations 
are reduced to single band ones. In this case the non-crossing approximation 
leads to a finite life time and the diagonal velocity vertex is renormalized 
after the summation of all non-crossing disorder paring lines. 

\begin{figure}[h]
\includegraphics[width=5cm]{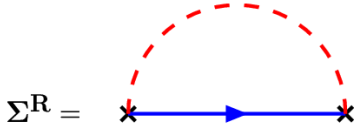}
\centering
\caption{Self-Energy diagram}
\label{selfen1}
\end{figure}

Only the imaginary part of the self-energy is important in the $\xi \ll 1$
limit.
The self-energy diagram that contributes to its diagonal element for the Gaussian disorder model is shown in Fig.~\ref{selfen1}. 
Because $G^{R-}(\epsilon)$ has a negligible imaginary part one can keep only the $G^{R+}(\epsilon)$ part of the Green's function in calculations of $\Sigma^R$, i.e.
$\Sigma^R=\sum_{{\bf k'}}V^{++}_{{\bf kk'}} G^{R+} V^{++}_{{\bf k'k}}$. Hence, the renormalized retarded Green's function is given
 by a standard result
\begin{equation}
G^{R+}_0(\epsilon)\rightarrow G^{R+}(\epsilon)=\frac{|u^{+}_{{\bf k}}\rangle \langle u^{+}_{{\bf k}}|}{\epsilon -\epsilon^{+}_{{\bf k}}+i/(2\tau^+)}
\label{rengr}
\end{equation}
where
\begin{equation}
1/\tau^+=-2 {\rm  Im}\left( \Sigma^R \right) = 2\pi\int \frac{d^2{\bf k'}}{(2\pi)^2} |V^{++}_{{\bf k'k}}|^2 \delta (\epsilon-\epsilon_{{\bf k'}}^+)
\label{taupl}
\end{equation}
Note that we have so far ignored the off-diagonal part of the self-energy which will contribute later 
when considering the diagrammatic expansion of the conductivity in powers of $\xi$.  It is convenient to organize
the calculation in this way because the off-diagonal parts do not need to be partially 
summed to infinite order and only a few terms contribute to the desired order.

\subsubsection{The vertex correction.}

Every disorder line that connects retarded and advanced Green's functions adds a factor $n_i V_0^2 \sim \xi$ multiplied by a product $G^R G^A$. 
While the former factor gives a power of $\xi$, the product of two Green's functions gives a $\xi^{-1}$ contribution if the pair of
Green's functions over which the momentum sum is performed are in the same band.
Thus one can approximate\cite{Inoue:2004_a}
\begin{equation}
G^{R+}(\epsilon)  G^{A+}(\epsilon) \approx 2 \pi \tau^+ \delta (\epsilon - \epsilon_{{\bf k}}).
\label{InoueGG}
\end{equation}
Since $\tau^+ $ is proportional to $1/\xi$ this product of Green's functions compensates the small factor due to disorder vertexes. 
Therefore terms $v_y^{++}$ and $\sum_{{\bf k'}} V^{++}_{{\bf kk'}}G^{R+} v_y^{++} G^{A+}V^{++}_{{\bf k'k}}$
are of the same order and one should perform the summation over all ladder diagrams that include only Green's functions in the positive energy
band.  These ladder diagrams simply renormalize the diagonal matrix element of the velocity vertex operator.

\begin{figure}
\includegraphics[width=7cm]{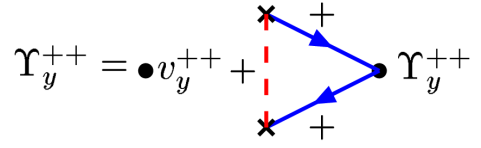}
\caption{Graphical representation of the vertex equation.}
\label{fvert2}
\end{figure}
The bare diagonal velocity is $v_y^{++}=v \sin(\theta)\sin(\phi)$. The vertex equation is
\begin{equation}
\Upsilon_y^{++}=v_y^{++}+ \int \frac{d^2{\bf k'}}{(2\pi)^2}  G^{A+} |V^{++}_{{\bf k'k}}|^2 G^{R+} \Upsilon_y^{++}
\label{veq2}
\end{equation}
Eq. (\ref{veq2}) for the vertex correction is represented graphically in Fig.~\ref{fvert2}.
This equation has a solution of the form $\Upsilon_y^{++} = \alpha v \sin(\theta) \sin(\phi)$ 
where $\alpha = \tau^{tr}/\tau^+$, 
\begin{equation}
\begin{array}{l}
1/\tau^{tr} =  2\pi\int \frac{d^2{\bf k'}}{(2\pi)^2} |V^{++}_{{\bf k'k}}|^2 (1-\cos({\bf k,k'})) 
 \delta (\epsilon_F-\epsilon_{{\bf k'}}^+)   \\
\\
=\frac{(vk_F)^2+4\Delta^2}{4\tau^q((vk_F)^2+\Delta^2)},
\end{array}
\label{tautr}
\end{equation}
\begin{equation}
1/\tau^q = \frac{nV_0^2k_F}{v_F},
\label{tauq}
\end{equation}
and $v_F$ is the Fermi velocity
 $v_F=(\partial \epsilon^+/\partial k)_{k=k_F}=v^2k_F/\epsilon_F$. 
Combining this result with Eq. (\ref{InoueGG})
gives the useful weak-scattering identity
 \begin{equation}
 G^{R+}_0 v_{x/y}^{++} G^{A+}_0 \rightarrow G^{R+} \Upsilon_{x/y}^{++} G^{A+} \approx v_{x/y}^{++} 2\pi \tau^{tr} \delta (\epsilon -\epsilon_{{\bf k}}^+). 
 \label{vyren}
 \end{equation} 

\subsubsection{Expansion in the number of virtual interband transitions.} 

After summing the series of diagrams that contain only diagonal disorder vertices 
one can look at the effect of the
inter-band disorder matrix elements $V^{+-}_{{\bf kk'}}$ and $V^{-+}_{{\bf kk'}}$. 
Any disorder average line with this component still adds a small factor
$n_i V_0^2$, however now it is followed by at least one Green's function in the lower band $G^{R-}$ or $G^{A-}$. 
The new Green's function products that accompany the interband disorder interaction lines, products like 
$G^{R-} G^{A+}$, {\em  etc.}, no longer can be considered as large. 
They are at most zeroth order in the parameter $\xi$ in the final expression rather than being proportional to $\xi^{-1}$. 
Since any part of a diagram containing either $G^{R-}$ or $G^{A-}$ adds a factor of the small parameter $\xi$, 
these diagrams do not have to be summed to infinite order.
It is enough to consider only the finite set of them that contribute at order $\xi^0$, 
the set of diagrams with only one $G^{R-}$ or $G^{A-}$ line.
The full set of such diagrams in the non-crossing approximation is shown in Fig.~\ref{mainfig} and in individual sets in
Figs.~\ref{fintr}, \ref{fsj}, \ref{fadist}, and \ref{om6}. Of course, to find the
total conductivity, one should consider them together with any number of intra-band disorder averaging lines which are already
included by replacing $v^{++}$ by $\Upsilon^{++}$.

\section{Identification of SBE and Kubo formalism AHE contributions.}
\label{comp}
After discussing the two formalisms we are now ready to demonstrate their equivalence and 
to identify the set of chiral basis diagrams that correspond to the various physical 
effects that have been identified in the more physically transparent semiclassical treatment.
It is important to note that the same result is obtained by doing the calculation
in the normal spin basis, as shown in Appendix~\ref{vertex_calculation}, but in that case
a given Feynman diagram can contain contributions from more than one of the 
effects identified in the semiclassical theory.
Calculations in the band eigenstate 
basis are necessary to interpret the SBE physics because the SBE approach starts 
by ignoring interband coherence and adds coherence effects by hand by positing side-jump
and anomalous velocity effects. 

\subsection{The intrinsic contribution}
\subsubsection{The intrinsic contribution from the SBE}
According to Eq. (\ref{jtotal_i}), to find the part of the Hall conductivity not related to impurity scattering one should first find the Berry curvature 
of both bands. In two dimensions, only the z-component of the Berry curvature is meaningful. From Eqs. (\ref{basis}) and (\ref{berrcurv}) for the Dirac Hamiltonian we find
\begin{equation}
F^{\pm}(k)=2{\rm Im} \langle \partial_{k_y} u_{{\bf k}}^{\pm} | \partial_{k_x} u_{{\bf k}}^{\pm} \rangle =
\mp \frac{\Delta v^2}{2(\Delta^2+(vk)^2)^{3/2}}
\label{Fpm}
\end{equation}
where $\pm$ in $F^{\pm}(k)$ refers to upper/lower bands.
Since the intrinsic contribution comes from the whole Fermi sea, both the totally filled negative energy and the partially filled 
positive energy bands contribute
to it. Substituting Eq.~(\ref{Fpm}) into Eq. (\ref{jtotal_i}) we find
\begin{equation}
\begin{array}{l}
\sigma_{xy}^{intr}=-e^2 \int_0^{\infty} \frac{d^2 {\bf k}}{(2\pi)^2} F^{-}(k)-\\
\\
-e^2\int_{k<k_F} \frac{d^2 {\bf k}} {(2\pi)^2} F^{+}(k)=-\frac{e^2\Delta } {4\pi \sqrt{\Delta^2+(vk_F)^2}}
\end{array}
\label{sxyintr1}
\end{equation}
which coincides with the first term in Eq. (\ref{sxy2}).

\subsubsection{The intrinsic contribution from the Kubo formalism}

\begin{figure}
\includegraphics[width=9cm]{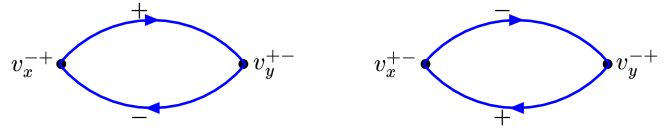}
\centering
\caption{Diagrams, without disorder paring lines, contributing to the Hall current. Plus/minus refer to the band index with positive/negative energy }
\label{fintr}
\end{figure}

Since the intrinsic contribution to the AHE, as identified in the semiclassical approach, 
does not depend on impurities, it is expected that this contribution must correspond
to the Kubo formula contribution from diagrams that do 
not involve any disorder lines, {\em i.e.} the bubble diagram conductivity with
all Green's functions bare.
In Fig.~\ref{fintr} we show the diagrams from $\sigma_{xy}^{I}$ 
that are non-zero even without disorder. 
The unperturbed Green's function is diagonal in the chiral basis given by Eq. (\ref{basis});
the band label in the diagram is 
"$+$" for $G_{0}^{R+/A+}$ and "$-$" for $G_{0}^{R-/A-}$.
The contribution to the Hall current due to these diagrams is
\begin{equation}
\begin{array}{l}
\sigma_{xy}^{I(int)}= \frac{e^2}{2\pi} \int \frac{d^2 {\bf k}}{(2\pi)^2} \frac{i\pi (v_x^{+-} v_y^{-+}-v_x^{-+}v_y^{+-})}{\epsilon_{{\bf k}}^+ -\epsilon_{{\bf k}}^-} 
\delta (\epsilon_F -\epsilon_{{\bf k}}^+ )=\\
\\
=-\frac{e^2\Delta } {4\pi \sqrt{\Delta^2+(vk_F)^2}}
\end{array}.
\label{sxyint}
\end{equation}

Surprisingly this result coincides with the semiclassical result obtained by evaluating Eq. (\ref{sxyintr1}) even
though we have not yet included the disorder free part of
$\sigma^{II}_{xy}$, which is also non-zero in general and must be included as a part of the intrinsic conductivity. 
It turns out however that for the case of the 2D Dirac model $\sigma^{II}_{xy}$ vanishes when the 
Fermi level is above of the gap. To see that we first observe that  
$\sigma^{II}_{xy}$ is exactly the sum of two contributions, one of which
is the same as the intrinsic conductivity given by the SBE Eq. (\ref{sxyintr1}) while 
the other is equal to the intrinsic part of $\sigma^{I}_{xy}$ but with the opposite sign:
\begin{equation}
\begin{array}{l}
\sigma^{II}_{xy} = -e^2 (\int_{\epsilon_{{\bf k}}^{+}<\epsilon_F} \frac{d^2 {\bf k}}{(2\pi)^2}  F^{+}(k) +\\
\\
+\int \frac{d^2 {\bf k}}{(2\pi)^2}  F^{-}(k) )  - \sigma^{I(int)}_{xy}
\end{array}
\label{sigmaIIfin1}
\end{equation}
Details of the calculation which lead to this result are presented in Appendix \ref{sigmaII_app}.
This result is of course not surprising when we recognize that Eq. (\ref{sxyintr1}) expresses
the linear response of a clean system, {\em i.e.} the full $\sigma_{xy}$ in the absence of disorder
under the influence of an accelerating electric
field; since $\sigma_{xy}$ must equal $\sigma_{xy}^I+\sigma_{xy}^{II}$ it follows that 
$\sigma_{xy}^{II}=\sigma_{xy}-\sigma_{xy}^I$. 
This observation also means that there is no special intrinsic contribution arising from the Fermi surface,
where occupation numbers change because of acceleration, because the 
corresponding terms in $\sigma^{I}_{xy}$ and $\sigma^{II}_{xy}$ cancel and the final result coincides with the 
semiclassical prediction Eq. (\ref{sxyintr1}), coming from all electrons {\em below} the Fermi surface.

We have already calculated the sum of the two integrals in Eq. (\ref{sigmaIIfin1}) and $\sigma^{I(int)}_{xy}$ (see Eqs. \ref{sxyintr1},
\ref{sxyint}) and showed that they have the same value.  It follows that in the metallic regime, when $\epsilon_F>\Delta$, 
$\sigma^{II}_{xy}$ is identically zero.
In the case of the 2D Dirac Hamiltonian, the Fermi surface contribution $\sigma^{I(int)}_{xy}=0$ 
vanishes when the chemical potential is in the gap ($-\Delta < \epsilon_F <\Delta$),  but
\begin{equation}
\sigma^{II(gap)}_{xy} = -e^2
\int \frac{d^2 {\bf k}}{(2\pi)^2} f(\epsilon_{{\bf k}}^{-}) F^{-}(k) = -\frac{e^2}{4\pi} \ne 0
\label{sigmaIIfin}
\end{equation}
(in units $\hbar=1$). 
Thus $\sigma^{II(gap)}_{xy}$ is responsible for the quantum Hall effect in the Dirac band.\cite{Mele:2004_a}

\subsection{Side-Jump related contributions.}

\subsubsection{Semiclassical side-jump conductivities.}

Since direct inter-band scattering is not energetically allowed in the weak disorder limit of our 
model, we consider only the coordinate shift effect of scattering in the
upper band:
\begin{equation}
 \delta {\bf r}_{{\bf k',k}} = \langle u_{{\bf k'}}^+| i\frac{\partial}{\partial {\bf k'}} u_{{\bf k'}}^+ \rangle - 
 \langle u_{{\bf k}}^+| i\frac{\partial}{\partial {\bf k}} u_{{\bf k}}^+ \rangle
 -  \hat{{\bf D}}_{{\bf k',k}} {\rm arg}(\langle u_{{\bf k'}}^+|u_{{\bf k}}^+\rangle )
\label{delr44}
\end{equation}
where ${\bf \hat{D}_{k',k}} = \frac{\partial }{\partial {\bf k'}} +  \frac{\partial }{\partial {\bf k}}$.
Evaluation of Eq.(~\ref{delr44}) for the positive energy band leads to the following expression after 
noting that the absolute magnitude of the momentum is conserved: 
\begin{equation}
\delta {\bf r}_{{\bf k',k}} = \frac{F^+(k) \; \hat{{\bf z}} \times ({\bf k'}-{\bf k})}{|\langle u_{{\bf k'}}^+|u_{{\bf k}}^+\rangle|^2}
\label{sj1}
\end{equation}
where  $F^+$ is the Berry curvature of the upper band, defined in Eq. (\ref{Fpm}).
The average side-jump velocity can be obtained by multiplying the transition rate and 
the side-jump associated with a particular transition as in Eq. (\ref{jtotal_sj}): 
\begin{equation}
\begin{array}{l}
 v_x^{sj}({{\bf k}}) =\sum_{\bf k'} \omega_{{\bf k'k}} (\delta {\bf r}_{{\bf k',k}})_x =\\
\\
2\pi n V_0^2 \int \frac{d^2{\bf k'}}{(2\pi)^2} (-F^+)(k_y^{\prime}-k_y) \delta (\epsilon_{{\bf k}}^+ -\epsilon^+_{{\bf k'}}) =F^+ k_y/\tau^q,
\end{array}
\label{vsj3}
\end{equation}
where we have used that in the lowest Born approximation $\omega_{{\bf k'k}}=2\pi |V_{{\bf k',k}}^{++}|^2 
\delta (\epsilon_{{\bf k}}^+ -\epsilon^+_{{\bf k'}})$.
The side-jump velocity does not produce any contribution to the total current in equilibrium, but in an external
electric field the nonequilibrium correction to the distribution function, Eq. (\ref{solg}), results in a non-zero 
net current. Disregarding a small antisymmetric part the distribution function correction is 
 $g_{{\bf k}}  = \left( - \partial f_{0}/\partial \epsilon_{{
\bf k}}^+ \right)eE_y (v_{0{\bf k}}^{++})_y \tau^{tr}$, where 
$(v_{0{\bf k}}^{++})_y = \partial \epsilon^+_{{\bf k}}/\partial k_y$.
It follows that the side-jump contribution to the SBE Hall conductivity is
\begin{equation}
\begin{array}{l}
\sigma_{xy}^{sj}=e\int \frac{d^2{\bf k}}{(2\pi)^2} (g_{{\bf k}}/E_y)  v_x^{sj}({{\bf k}}) = \\
\\
=-\frac{e^2\Delta (vk_F)^2}{2\pi 
\sqrt{(vk_F)^2 +\Delta^2} ((vk_F)^2+4\Delta^2)}.
\end{array}
\label{vsj4}
\end{equation}

The second side-jump effect discussed in Section III follows from the change of energy
of the scattered particle after the coordinate shift at a scattering event in the 
presence of an external electric field. It results in the anomalous correction to the distribution function that can be determined from
Eq.(\ref{ganl}).  In our case
\begin{equation}
\sum_{{\bf k'}}\omega_{{\bf k,k'}} 
 \left( g^{adist}_{{\bf k}} - g^{adist}_{{\bf k'}} +\frac{-\partial f_0}{\partial \epsilon_{{
\bf k}}^+} e E_y (\delta r_{{\bf k',k}})_y \right) =0.
\label{ganlb}
\end{equation}
Substituting $\delta r_{{\bf k',k}}$ from Eq. (\ref{sj1}) and looking for a solution of Eq. (\ref{ganlb}) in the form 
$g^{adist}_{{\bf k}}=\gamma k_x$ we find that $\gamma=
F^+\frac{-\partial f_0}{\partial \epsilon_{{\bf k}}^+} e E_y\tau^{tr}/\tau^q$.
The corresponding contribution to the Hall conductivity is
\begin{equation}
\begin{array}{l}
\sigma_{xy}^{adist}=e\int \frac{d^2{\bf k}}{(2\pi)^2} (g_{{\bf k}}^{adist}/E_y)  v_x^{++} = \\
\\
=-\frac{e^2\Delta (vk_F)^2}{2\pi 
\sqrt{(vk_F)^2 +\Delta^2} \; ((vk_F)^2+4\Delta^2)}
\end{array}
\label{vsj44}
\end{equation}
For the 2D Dirac Hamiltonian we find that $\sigma_{xy}^{adist}=\sigma_{xy}^{sj}$, in an agreement with Ref. \onlinecite{Sinitsyn:2005_a} and \onlinecite{Sinitsyn:2005_b}, where
we showed that in the absence of direct interband scatterings these contributions must be identical. 
Thus the total side-jump related conductivity is just twice the one in Eq. (\ref{vsj44}):
\begin{equation}
\sigma_{xy}^{sj}+\sigma_{xy}^{adist}=-\frac{e^2 \Delta (vk_F)^2}{\pi 
\sqrt{(vk_F)^2 +\Delta^2} ((vk_F)^2+4\Delta^2)}.
\label{vsj4}
\end{equation}
This result coincides with the second term in the total conductivity Eq. (\ref{sxy2}).
\begin{figure}
\includegraphics[width=9cm]{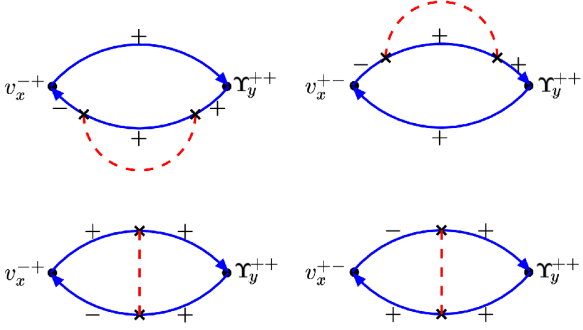}
\centering
\caption{Diagrams contributing to the side-jump current}
\label{fsj}
\end{figure}

\begin{figure}
\includegraphics[width=9cm]{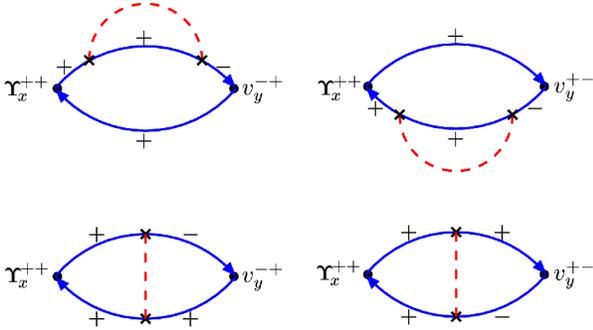}
\centering
\caption{Diagrams contributing to the current that arises from the anomalous distribution. High symmetry with side-jump diagrams in Fig.\ref{fsj} explains
the equality of two contributions to Hall current.}
\label{fadist}
\end{figure}
\subsubsection{Side jump effects in the Kubo formalism.}

If we look at the expression for the conductivity $\sigma^{Ia}_{xy} \sim {\rm Tr}(\hat{v}_x G^R \hat{v}_y G^A)$, we find two velocity operators. Tracing the history
of how
this expression was derived, the operator $\hat{v}_x$ appeared when we were looking for its average $\langle \hat{v}_x\rangle$ and the operator $\hat{v}_y$
appeared from the coupling of the charge 
current to the electric field. Thus the right velocity vertex in Feynman diagrams is associated with the immediate
effect of the electric field, such as the acceleration of wave packets (due to the coupling of electric field  to the 
diagonal velocity) or the mixing  of states from different bands (due to 
the coupling to the off-diagonal velocity).
In turn, the left velocity vertex shows what average velocity is finally induced due to this effect if this diagram is nonzero.

This observation allows us to look for the diagrams responsible for various contributions to the AHE.  
The side jump current is due to the accumulation of transverse
coordinate shifts. These shifts are anomalous, {\em i.e.} they 
arise from inter-band coherence induced by impurities.
 In the operator language, 
they appear because the expectation value of the off diagonal 
part of the velocity operator becomes nonzero on average. Hence 
the diagrams responsible for the side jump effect are expected to have 
off diagonal elements $v_x^{+-}$ or $v_x^{-+}$ on their left side. 
In the SBE the side jump velocity becomes nonzero on
average only after the electric field induces
the nonequilibrium correction to the distribution function $g_{{\bf k}}$. That correction is due to the coupling of the electric field to the 
diagonal part of the velocity. Hence side jump diagrams should have only diagonal velocity $v_y^{++}$ on their right hand side.
There are four different diagrams, shown in Fig.~\ref{fsj} with such properties. 
Their calculation, using Eq. (\ref{rengr}) and (\ref{vyren}), leads to the same result as the one in Eq. (\ref{vsj3}).
The band labels at the right vertex in these diagrams are near the Fermi 
energy in the occupied band. 

The identification of the diagrams which contain the contribution obtained in the SBE 
due to the anomalous distribution that arises from the side-jump, $\sigma_{xy}^{adis}$, can
be identified by similar reasoning.  For these contributions, the electric field 
is coupled to the coordinate shift (and hence
to the off-diagonal velocity). The coupling changed the band energy upon scattering,
leading to the anomalous  correction to the distribution function. After this
the average of the diagonal velocity from such a distribution correction becomes nonzero.
Hence the corresponding diagrams should have the off diagonal
velocity on the right and the diagonal velocity on the left. 
There are again four such diagrams in Fig.~\ref{fadist} whose evaluation leads to the same result as Eq. (\ref{vsj44}). 
Symmetry relations between the Feynman diagrams  
shown in Fig.~\ref{fsj} and in Fig.~\ref{fadist} explain why the two 
contributions are equal to each other.
However, within the semiclassical formalism, the seemingly exact relation between the two contributions is not clear to us. 

We emphasize once again that here only four diagrams for each of the two
contribution have been evaluated to obtain the result of order $O(\xi^{0})$ and not a partial infinite sum.
(This counting assumes that $\Upsilon^{++}$ has already been evaluated separately.) 
We also emphasize that although the explicit demonstration of equivalence between
these specific Feynman diagrams and the side-jump calculations which we have identified on 
physical grounds is for the Dirac model alone, the considerations
on which the identification has been made is general, and we believe that the 
identification is generally valid. 

\subsection{Skew scattering}
\label{skew_section}
\subsubsection{Skew scattering in the SBE.}
\begin{figure}
\includegraphics[width=5cm]{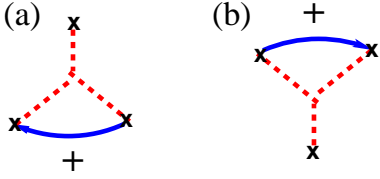}
\centering
\caption{Lowest order non-zero terms in the Born series responsible for the asymmetric part of $|T_{{\bf kk'}}|^2$.}
\label{fskew3T} 
\end{figure}

The skew scattering contribution originates microscopically from asymmetry in the scattering rate. 
This asymmetry can be estimated for known disorder scattering sources 
by truncating the Born series expansion of the $T$-matrix at low order. 
As discussed in Sec. \ref{SBE}, in the lowest Born approximation the scattering rate 
is symmetric with respect to an exchange of indexes. The first 
non-zero contribution appears at the next (3rd) order. 
In Eq. (\ref{om3}) the terms in parentheses represent the absolute value of the squared $T$-matrix element
$|T_{{\bf k,k'}}|^2$ to the order $V^3$. In Fig.~\ref{fskew3T} we show the diagrams representing both such terms from Eq. (\ref{om3}).  
This contribution vanishes for a Gaussian (white noise) disorder distribution. 
However, one should generally keep this contribution, even if $V_1$ is small, because it leads to a parametrically different 
and hence experimentally distinguishable contribution to the AHE. 

\begin{figure}
\includegraphics[width=6.5cm]{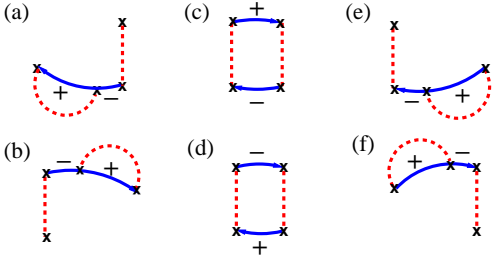}
\centering
\caption{Terms contributing to the antisymmetric part of $|T_{{\bf k'k}}|^2$ in
the white noise non-crossing approximation
at 4th order in the disorder potential.}
\label{fskew4T}
\end{figure}

At 4-th order in $V$ the number of terms that contribute to the asymmetric scattering rate,
and hence skew scattering, is non-zero even for Gaussian disorder models.
In Fig.~\ref{fskew4T} we show schematically all the terms that contribute
at 4-th order in the non-crossing white noise approximation. For example, the diagram in
Fig.~\ref{fskew4T}(a) corresponds to the expression 
$$
\sum_{{\bf k''}}
\langle V_{{\bf kk'}}^{++} V_{{\bf k'k}}^{+-}\rangle_{dis}
 \frac{1}{\epsilon-\epsilon^-_{{\bf k}}-i\eta} 
\langle V_{{\bf kk''}}^{-+} V_{{\bf k'' k}}^{++} \rangle_{dis}
 \frac{1}{\epsilon-\epsilon_{{\bf k''}}^{+} -i\eta}. 
$$
The summation of terms in Figs.~\ref{fskew3T} and \ref{fskew4T} leads to the following expression for the asymmetric part of the scattering rate:
\begin{equation}
\omega^{(a)}_{{\bf k,k'}} \approx \omega^{(3a)}_{{\bf k,k'}} +\omega^{(4a)}_{{\bf k,k'}}
\label{om34}
\end{equation}
where
\begin{equation}
\omega^{(3a)}_{{\bf k,k'}}  = \frac{\pi n_i V_1^3 \Delta k^2 } {2[(vk)^2+\Delta^2]} \sin(\phi-\phi')\delta (\epsilon_{{\bf k}}^+ -\epsilon_{{\bf k'}}^+) 
\label{om33}
\end{equation}

\begin{equation}
\omega^{(4a)}_{{\bf k,k'}}=\frac{3 \pi\Delta k^2 (n_i V_0^2)^2} {4[(vk)^2+\Delta^2]^{3/2}} \sin(\phi-\phi')\delta(\epsilon^+_{{\bf k}}-
\epsilon^+_{{\bf k'}}).
\label{om44}
\end{equation}
The dependence of $\omega^{(3a)}_{{\bf k,k'}}$ and $\omega^{(4a)}_{{\bf k,k'}}$ on the impurity concentration $n_i$ provides valuable information on
the physics behind asymmetric scattering. Although $\omega^{(3a)}_{{\bf k,k'}}$ is proportional to $n_i$ and is therefore due to separate incoherent contributions from all impurities, the 4th order Gaussian contribution $\omega^{(4a)}_{{\bf k,k'}}$ is proportional to $n_i^2$ and can 
be interpreted as due to interference between different scattering centers.
Given the scattering rate one can calculate the transverse scattering time and the Hall current 
using Eqs. (\ref{taus}) and (\ref{cond1}).
The skew scattering contribution to the Hall conductivity is then 
\begin{equation}
\begin{array}{l}
\sigma_{xy}^{sk}=
-\frac{e^2 3\Delta (vk_F)^4}{4\pi \sqrt{(vk_F)^2+\Delta^2} [4\Delta^2+(vk_F)^2]^2} 
- \frac{ e^2V_1^3 }{2\pi n_i V_0^4 } \frac{\ (vk_F)^4 \Delta}{  (4\Delta^2+(vk_F)^2)^2} 
\end{array}
\label{jt}
\end{equation}
This result coincides with the last two terms in Eq. (\ref{sxy2}). Note also that the second term in Eq. (\ref{jt})
is inversely proportional to the impurity concentration $n_i$ and thus the transverse conductivity due to this skew scattering
becomes large in the clean limit whenever $V_1\ne 0$. 
In contrast, the first term in Eq. (\ref{jt}) appears even for the Gaussian disorder correlations and it is independent of the 
impurity concentration. This is a result that has not been previously noticed: \emph{the skew scattering contribution, as defined 
through the collision term in the SBE, can give
a result independent of the impurity concentration and disorder strength}.  
Of course this contribution is subdominant if disorder is weak. 
\begin{figure}
\includegraphics[width=9cm]{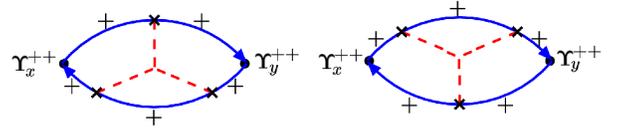}
\caption{Diagrams contributing to 3rd order skew scattering.}
\label{fskew}
\end{figure}
\begin{figure}
\includegraphics[width=9cm]{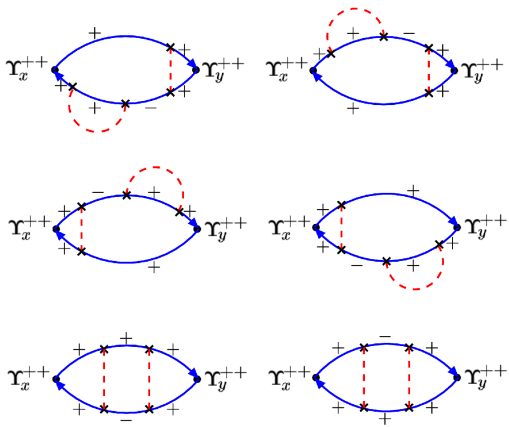}
\caption{Leading diagrams in self-consistent non-crossing approximation for the
skew scattering contribution 
to the AHE in the semiclassical approach.}
\label{om6} 
\end{figure}
\subsubsection{Skew scattering in the Kubo formalism.}

We have shown above that intrinsic and side-jump related diagrams contain at least one 
off-diagonal velocity operator element, $v^{\pm}$. 
Skew scattering is, from this conceptual point of view, the most conventional 
contribution to the AHE since it depends only on velocity operator matrix elements that are 
diagonal in the band indices, $v_{x/y}^{++}$, or more accurately its renormalized version $\Upsilon_{x/y}^{++}$. 
In Figs.~\ref{fskew} and \ref{om6} we show respectively the leading order diagrams for non-Gaussian 
correlations and for Gaussian disorder in the non-crossing approximation.

The correspondence between calculations of skew scattering contributions based on the Kubo formalism and the 
SBE becomes obvious when we compare Figs.~\ref{fskew} and \ref{om6} with
terms contributing to the asymmetric scattering rate in Figs.~\ref{fskew3T} and \ref{fskew4T}. 
Comparing both sets of figures shows that the central part in 
Figs.~ \ref{fskew} and \ref{om6} coincides with the formal perturbation expansion of the asymmetric part of $|T_{{\bf k,k'}}|^2$.
Evaluating the diagrams in Figs.~\ref{om6} and ~\ref{fskew}, using the expressions for the renormalized
diagonal velocities $\Upsilon^{++}_{x/y}$ given by Eq. (\ref{vyren}),
one obtains the first and second terms in Eq. (\ref{jt}) respectively. 

This final result completes our comparison showing the equivalence between the SBE and the Kubo formalism.

\section{Excluded diagrams}
\label{missing}
\begin{figure}
\includegraphics[width=4.4cm]{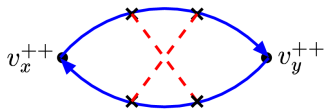}
\caption{A diagram with crossing disorder lines.}            
\label{fcross} 
\end{figure}

In this paper we have considered only diagrams without crossing disorder paring lines.
This approximation is self-consistent and justified by Ward identities but
incomplete. There is no obvious reason why crossing diagrams like the one in Fig.~\ref{fcross} are small
in comparison with others. 
It is expected, however, that crossing diagrams
are parametrically different from others and thus the self-consistent non-crossing approximation can be justified in many cases. 

\begin{figure}[h]
\includegraphics[width=5.0cm]{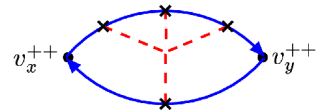}
\caption{A higher order diagram contributing to the skew scattering current. }
\label{fskew4}
\end{figure}

Another type of terms that we dropped from our discussion is due to higher order non-Gaussian correlations.
An example is shown in Fig.~\ref{fskew4}.  This diagram leads to current that behaves as $1/n_i$ on the impurity concentration. 
Although formally it is small in comparison with similar diagrams from Fig.~\ref{fskew}, when many different types of impurities are involved, the 3rd cumulant of the disorder potential can be suppressed because different types of impurities may contribute to it with different signs. In contrast, the
4th order cumulants remain of the same sign and thus can compete with the 3rd one.
We are not aware of any work on crossing diagrams and the 4th order 
non-Gaussian skew scattering contributing to the AHE.

\section{Conclusion}
\label{conclusion}
We have compared two different techniques commonly used for anomalous Hall current calculations. 
One is based on the semiclassical Boltzmann equation (SBE) and on the heuristic 
identification of side-jump and anomalous velocity unconventional contributions.
The second approach is systematic linear response theory, which requires evaluation of 
a Kubo formula.  The SBE lacks a systematic derivation in a controlled expansion.
In particular, it is not completely obvious that all unconventional effects 
have been identified or that these effects represent independent contributions that 
should be added together.  It is based on the assumption that electrons in the diffusive regime
can be treated as classical particles and only  the collision term and the expression
for the velocity need to be calculated quantum mechanically. This assumption fails,
for example, when localization effects are important.
The semiclassical approach, however, has several advantages, such as 
the attractive transparency of its physical interpretation 
and its applicability
 beyond the linear response approximation. 
It is well known that in the theory of the longitudinal 
conductivity, the SBE is equivalent to the self-consistent 
non-crossing approximation in the Kubo formalism.  
We have extended this correspondence between SBE and Kubo
approaches by identifying the Feynman diagrams that correspond
to the unconventional side-jump and anomalous velocity effects 
that are important for the anomalous Hall effect. 
We have established a one to one correspondence between both 
techniques in leading order of the small parameter $\xi=1/k_F l_{sc}$. 
This direct correspondence between particular diagrams 
and various side-jump and 
anomalous velocity effects applies only when working in the Kubo formalism with
the chiral exact Bloch eigenstate representation.
The identification is based in part on the physical
transparency of the SBE AHE contributions, has been verified 
by explicit calculation for the 2D Dirac model, and is completely general. 
This connection between the two approaches has not been directly established 
in previous literature, to our knowledge,
which contains many farraginous results which often 
appear to be inconsistent with each other. 
  
In addition, in performing this comparison with the skew scattering contribution,
we have discovered that there is a contribution that 
semiclassically is also related to the asymmetric part of the collision term kernel, but which leads 
 to a contribution independent on the impurity concentration and is non-zero even for Gaussian correlations.
 The identification of an impurity-independent skew scattering contribution has not been 
made in previous literature.
Although this contribution to the Hall current is independent of impurity
 concentration, like the side-jump and intrinsic contributions, 
it has a distinct physical meaning and thus has 
 generally different dependence on other parameters, e.g. $\epsilon_F$, as one can see from our results
for the 2D Dirac Hamiltonian.
 
The Kubo formalism and the SBE should be considered as complementary techniques. 
The former provides a rigorous justification for the latter.  The latter
provides simple semiclassical interpretations of the predictions of the former,
 which is important in developing intuition that can be applied to more complicated problems. 
We demonstrated in this work for the first time how very different approaches of the Hall current calculation converge to the 
same expressions.  We hope that the observations made in this paper will help 
increase the consistency of various publications 
devoted to disorder contributions in the AHE.
 
\begin{acknowledgments}
The authors greatly acknowledge insightful conversations with  J. Inoue, T. Nunner, N. Nagaosa, S. Onoda, and G. Bauer.
This work was supported by the NSF under the grant DMR-0547875, by the ONR under grant
ONR-N000140610122, by the KITP NSF grant PHY99-07949, by grant No. POCI/FIS/58746/2004
in Portugal, by the Polish State Committee for Scientific Research under Grant No. 2~P03B~053~25, by STCU Grant No.~3098
in Ukraine, by the Welch Foundation and by DOE grant DE-FG03-02ER45958.  Jairo Sinova is a Research Corporation Cottrell Scholar.
\end{acknowledgments}

\appendix

\section{Vertex correction calculation}
\label{vertex_calculation}
In this appendix we discuss the procedure to calculate the vertex correction to the velocity operator in the presence
of disorder in the normal spin basis. This ladder vertex corrections is defined by 
$\tilde{v}_\alpha(\epsilon_F)\equiv v_\alpha+\delta\tilde{v}_\alpha(\epsilon_F)$
\begin{eqnarray}
\delta\tilde{v}_\alpha(\epsilon_F)&=&\frac{n_i V_0^2}{V}\sum_\kk G(\epsilon_F) J_\alpha G(\epsilon_F)\\
\nonumber&&+\frac{n_i V_0^2}{V}\sum_\kk G^R(\epsilon_F) \delta\tilde{J}_\alpha(\epsilon_F) G^A(\epsilon_F)
\end{eqnarray}
We decompose the operators in the Pauli matrix representation 
$\delta\tilde{v}_\alpha=\delta\tilde{v}^\alpha_0\sigma_0+\delta\tilde{v}^\alpha_i \sigma_i$,
$G=G_0\sigma_0+G_i\sigma_i$, $v_\alpha=v^\alpha_0\sigma_0+v^\alpha_i \sigma_i$, where the coefficients can be read
from the expression for $G$:
\begin{equation}
\begin{array}{l}
G^R=\frac{1}{1/G_0^R - \Sigma^{R}}
=\frac{\epsilon_F+i\Gamma +v(k_x \sigma_x +k_y \sigma_y)+(\Delta - i \Gamma_1) \sigma_z}
{(\epsilon_F-\epsilon^+ +i\Gamma^+)(\epsilon_F -\epsilon^- +i\Gamma^-)}
\end{array}
\label{gf2}
\end{equation}
where $\Gamma=\pi n_iV_0^2/(4 v^2)$, $\Gamma_1 = \Gamma \cos(\theta)$, $\gamma^\pm =\Gamma_0 \pm\Gamma_1 \cos(\theta)$,
and $\cos\theta=\Delta/\sqrt{(vk)^2+\Delta^2}$. 

 We can write a closed equation for the vector $\delta\tilde{v}$ after multiplying the equation by $\sigma_i$ from the left and taking the trace:

\begin{eqnarray}
2\delta\tilde{v}_\alpha=C^\alpha+A\delta\tilde{v}_\alpha\\
\delta\tilde{v}_\alpha=(2-A)^{-1}C^\alpha
\label{deltav}
\end{eqnarray}
where the vector $C^\alpha=(C^\alpha_0,C^\alpha_x,C^\alpha_y,C^\alpha_z)$ and matrix $A$ are
defined as
\begin{equation}
C^\alpha_{a}\equiv \frac{n_i V_0^2}{V}\sum_\kk {\rm Tr}[\sigma_a G J_\alpha G],\,\,\,\,
A_{ab}\equiv \frac{n_i V_0^2}{V}\sum_\kk {\rm Tr}[\sigma_a G \sigma_b G]
\label{A_C}
\end{equation}
Here $a,b=0$ through $3$ and we reserve $i,j$ to label $x,y,z$.
Simplifications of these expressions can be obtained through the use of the following relations
\begin{eqnarray}
{\rm Tr}[\sigma_i\sigma_{i'}\sigma_j\sigma_{j'}]&=&
{\rm Tr}[(\delta_{ii'}+i\epsilon_{ii'k}\sigma_k)(\delta_{jj'}+i\epsilon_{jj'k'}\sigma_k')]\nonumber\\&&
=2(\delta_{ii'}\delta_{jj'}-\delta_{ij}\delta_{j'i'}+\delta_{ij'}\delta_{ji'}),
\end{eqnarray}
\begin{equation}
{\rm Tr}[\sigma_i\sigma_{j}\sigma_j']={\rm Tr}[i\epsilon_{ijk}\sigma_k \sigma_{j'}]=
2i\epsilon_{ijj'}.
\end{equation}



For the case of the Driac model Hamiltonian $v^y_y=v$ and $v^y_x=v^y_z=v^y_0=0$. Also $G_{x/y}\propto k_{x/y}$ and
$G_{0/z}$ do not have any angular dependence. The only non-zero terms in $A$ and $C$ are:
\begin{eqnarray}
C_x^y&=&
-\frac{4n_i V_0^2 v}{V}\sum_{\kk}{\rm Im}[G_0^{\rm ret}G_z^{\rm adv}]\equiv v b\nonumber\\
C_y^y&=&
 \frac{2n_i V_0^2 }{V}\sum_{\kk}v(G_0^{\rm ret}G_0^{\rm adv}-G_z^{\rm ret}G_z^{\rm adv})\equiv v a\nonumber\\
\nonumber\\
A_{00}&=&\frac{n_i V_0^2}{V}\sum_{\kk} (G_+G_++G_-G_-)\nonumber\\
A_{z0}&=&A_{0z}=
\frac{4n_i V_02}{V}\sum_{\kk}{\rm Re}[G_zG_0]\nonumber\\
A_{xx}&=&A_{yy}=va
\nonumber\\
A_{zz}&=&\frac{2n_i V_0^2}{V}\sum_{\kk}(G_0G_0+G_zG_z-G_xG_x-G_xG_x)\nonumber\\
A_{xy}&=&-A_{yx}=vb
\nonumber\\
\end{eqnarray}
This gives a block diagonal form to the matrix A so in what follows and 
since $C^y_0=C_z^y=0$ we will ignore the $A_{00},A_{zz},A_{z0},A_{0z}$ terms since irrespective of their value
$\delta\tilde{v}^y_0=\delta\tilde{v}^y_z=0$. The main step we have left to do is to compute $a$ and $b$
defined above:
\begin{eqnarray}
a&\equiv&\frac{2n_i V_0^2 }{V}\sum_{\kk}v(G_0^{\rm ret}G_0^{\rm adv}-G_z^{\rm ret}G_z^{\rm adv})\nonumber\\&=&
\frac{2n_i V_0^2 }{V}\sum_{\kk}\frac{\epsilon_F^2+\Gamma^2-h^2-\tilde{\Gamma}^2}{D_+D-}\\&=&
{2n_i V_0^2 }\left(\frac{\pi\nu_+}{\Gamma_+(4\lambda_+^2+\Gamma_-^2)}+
\frac{\pi\nu_-}{\Gamma_-(4\lambda_+^2+\Gamma_+^2)}\right)\nonumber\\&&\times
(\epsilon_F^2+\Gamma^2-h^2-\tilde{\Gamma}^2),
\end{eqnarray}
where we have defined $D_{\pm}\equiv((\epsilon_F-E_\pm)^2+\Gamma_\pm^2)$. 
Because $\lambda_+^2=\lambda_-^2=\epsilon_F^2\equiv \lambda$ and since 
$\Gamma_\pm=\Gamma(1\pm\cos^2\theta)$ for $\epsilon_F<-h$ 
and  $\Gamma_\pm=\Gamma(1\mp\cos^2\theta)$ for $\epsilon_F>h$ we can write this using the fact that
$2 \Gamma/\pi= n_i V_0^2(\nu_++\nu_-)$:
\begin{eqnarray}
a&=&4 \Gamma
\frac{(\epsilon_F^2+\Gamma^2-h^2-\tilde{\Gamma}^2)}
{(1+\cos^2\theta)(4\lambda^2+\Gamma2\sin^2\theta)}.\nonumber\\
&=&\frac{\sin^2\theta}{(1+\cos^2\theta)}+O(\frac{\Gamma^2}{\epsilon_F^2})\,\,\,{\rm for}\, |\epsilon_F|>h
\end{eqnarray}
For $b$ we have 
\begin{eqnarray}
b&=&-\frac{4n_i V_0^2 }{V}\sum_{\kk}{\rm Im}[G_0^{\rm ret}G_z^{\rm adv}]=
-\frac{4n_i V_0^2 }{V}\sum_{\kk} \frac{-\Gamma h+\epsilon_F\tilde\Gamma}{D_+D_-}\nonumber\\
&=&-4n_i V_0^2 \left(\frac{\pi\nu_+}{\Gamma_+(4\lambda_+^2+\Gamma_-^2)}+\right. \nonumber\\&&\left.
\frac{\pi\nu_-}{\Gamma_-(4\lambda_+^2+\Gamma_+^2)}\right)(-\Gamma h+\epsilon_F\tilde\Gamma)\nonumber\\
&=&\frac{4 \Gamma cos\theta}{(1+\cos^2\theta)\lambda}+\Gamma O(\frac{\Gamma^2}{\epsilon_F^2})\,\,\,{\rm for}\, |\epsilon_F|>h
\end{eqnarray}
where we have used the combination $\epsilon_F \tilde\Gamma=-\lambda\Gamma\cos\theta$. The matrix A reads (the quadrant of x and y):
\begin{equation}
A=\left(
\begin{array}{cc}
a&b\\
-b&a
\end{array}\right)
\,\,{\rm and}\,\, 
C=\left(
\begin{array}{c}
b\\
a
\end{array}\right)
\end{equation}
and putting this back in Eq. \ref{deltav} we finally get the renormalized vertex:
\begin{equation}
\tilde{v}_y=v8\Gamma \cos\theta\frac{(1+\cos^2\theta)}{\lambda(1+3\cos^2\theta)^2}\sigma_x
+\left(v+v\frac{\sin^2\theta}{(1+3\cos^2\theta)}\right)\sigma_y.
\end{equation}

\section{Derivation of the off diagonal conductivity using causal Green's functions.}
\label{Dugaev}

The off-diagonal DC conductivity can be calculated 
 using the Kubo formula representation in terms of
 causal Green's functions in the
limit $\omega \to 0$. We start from
\begin{eqnarray}
\label{C1}
\sigma _{xy}(\omega )
=\frac{e^2}{\omega }\,
{\rm Tr}\int \frac{d\varepsilon }{2\pi }
\frac{d^2{\bf k}}{(2\pi )^2}\;
\Upsilon _x\, G(\varepsilon +\omega )\, v_y\, G(\varepsilon )
\end{eqnarray}
with the Green's functions which include a self energy related
to the scattering from impurities.

The result of integration over $\varepsilon $ can be presented
in the following form
\begin{equation}
\begin{array}{l}
\sigma _{xy}(\omega )=\frac{ie^2}{\omega }
{\rm Tr}\int \frac{d^2{\bf k}}{(2\pi )^2}
\{ \\
\\
 f(-\epsilon ) v_x 
\frac{-\epsilon +v\bsig \cdot {\bf k}+\Delta \sigma _z}
{-2\epsilon }
v_y
\frac{-\epsilon -\omega +v\bsig \cdot {\bf k}+\Delta \sigma _z}
{(-\omega )(-2\epsilon -\omega )} \\
\\
+f(-\epsilon ) v_x
\frac{-\epsilon +\omega +v\bsig \cdot {\bf k}+\Delta \sigma _z}
{\omega (-2\epsilon +\omega)}
v_y
\frac{-\epsilon +v\bsig \cdot {\bf k}+\Delta \sigma _z}
{-2\epsilon } \\
\\
+f(\epsilon ) v_x
\frac{\epsilon +v\bsig \cdot {\bf k}+\Delta \sigma _z}{2\epsilon }
v_y
\frac{\epsilon -\omega +v\bsig \cdot {\bf k}+\Delta \sigma _z}
{(-\omega )(2\epsilon -\omega )} \\
\\
+f(\epsilon ) v_x
\frac{\epsilon +\omega +v\bsig \cdot {\bf k}+\Delta \sigma _z}
{\omega (2\epsilon +\omega )}
v_y
\frac{\epsilon +v\bsig \cdot {\bf k}+\Delta \sigma _z}
{2\epsilon }\\
\\
+\left[ f(\epsilon +\omega )-f(\epsilon ))\right]  v_x
\frac{\epsilon +\omega +v\bsig \cdot {\bf k}+\Delta \sigma _z-i\Gamma _0+i\Gamma _++i\Gamma _1\sigma _z}
{\omega (2\epsilon +\omega -i\Gamma _-+i\Gamma _+)}
v_y\\
\\
\,\,\,\,\,\,\,\,\,\,\,\,\,\,\,\,\,\,\,\,\,\,\,\,\,\,\,\,\,\,\,\,
\frac{\epsilon +v\bsig \cdot {\bf k}+\Delta \sigma _z-i\Gamma _0+i\Gamma _++i\Gamma _1\sigma _z}
{2\epsilon -i\Gamma _-+i\Gamma _+}\\
\\
+\left[ f(\epsilon )-f(\epsilon +\omega ))\right] \Upsilon _x
\frac{\epsilon +\omega +v\bsig \cdot {\bf k}+\Delta \sigma _z+i\Gamma _0+i\Gamma _+-i\Gamma _1\sigma _z}
{(\omega +2i\Gamma _+)(2\epsilon +\omega +i\Gamma _-+i\Gamma _+)}
v_y \\
\\
\,\,\,\,\,\,\,\,\,\,\,\,\,\,\,\,\,\,\,\,\,\,\,\,\,\,\,\,\,\,\,\,
\frac{\epsilon +v\bsig \cdot {\bf k}+\Delta \sigma _z-i\Gamma _0+i\Gamma _++i\Gamma _1\sigma _z}
{2\epsilon -i\Gamma _-+i\Gamma _+} \}
\end{array}
\label{C2}
\end{equation}
where $\epsilon \equiv \epsilon ^+_{\bf k}$,
Note that the renormalization of the vertex $v_x\to \Upsilon _x$
is included only in the last term. Even though the 5th term
contains the integration near the Fermi level, it should include
the unrenormalized
vertex. This is because the corresponding Green's functions come
with different signs of $\varepsilon +\omega $ and $\varepsilon $
only in the last term.
For definiteness, we assumed that the Fermi level is located in the upper band,
$\epsilon _F>\Delta $.

The first four terms in (\ref{C2}) correspond to the contribution of
states below the Fermi energy, $\sigma _{xy}^{int}$, which does
not have the vertex
corrections and is not affected by impurities. The 1st and 2nd
terms are associated with the filled lower Dirac band, whereas the 3rd and
the 4th are due to the upper band.
In the limit of $\omega \to 0$ it gives
\begin{equation}
\label{C3}
\sigma _{xy}^{int}=\frac{e^2 v^2\Delta }{2}
\int \frac{d^2{\bf k}}{(2\pi )^2}
\frac{f(\epsilon )-f(-\epsilon )}{\epsilon ^3}
\end{equation}

The 5th and 6th terms in (\ref{C2}) give two contributions to $\sigma _{xy}$.
We can identify them as $\sigma _{xy}^{b}$ and $\sigma _{xy}^{a}$,
respectively, because they are related to integrals of
$G^AG^A$ (i.e., with $\varepsilon +\omega <0,\, \varepsilon <0$ in
a corresponding product of causal functions) and $G^RG^A$
[i.e., with $\varepsilon +\omega >0,\, \varepsilon <0$ in
$G(\varepsilon +\omega )\, G(\varepsilon )$].

For $\omega ,\Gamma \to 0$ we find from (\ref{C2})
\begin{equation}
\label{C4}
\begin{array}{l}
\sigma ^a_{xy}(\omega )+\sigma ^b_{xy}(\omega )=ie^2
{\rm Tr}\int \frac{d^2{\bf k}}{(2\pi )^2}
\left( -\frac{\partial f}{\partial \epsilon }\right)
[  \\
\\
-v\sigma _x
\frac{\epsilon +\omega +v\bsig \cdot {\bf k}+\Delta \sigma _z}
{2\omega \epsilon }
v\sigma _y
\frac{\epsilon +v\bsig \cdot {\bf k}+\Delta \sigma _z}
{2\epsilon }\\
\\
+\Upsilon _x
\frac{\epsilon +\omega +v\bsig \cdot {\bf k}+\Delta \sigma _z+i\Gamma _0-i\Gamma _1\sigma _z}
{2(\omega +2i\Gamma _+)\epsilon }\,
v\sigma _y\\
\\
\,\,\,\,\,\,\,\,\,\,\,\,\,\,\,\,\,\,\,\,\,\,\,\,\,\,\,\,\,\,\,\,\,\,\,\,\,\,\,
\frac{\epsilon +v\bsig \cdot {\bf k}+\Delta \sigma _z-i\Gamma _0+\Gamma _1\sigma _z}
{2\epsilon }]
\end{array}
\end{equation}
As we see, the second term in (\ref{C4}) gives different results in the limits of
$\omega \ll \Gamma _+$ and $\omega \gg \Gamma _+$ (pure and "dirty" cases).
In the pure case we should also take $\Gamma _x\to v\sigma _x$.
Then, as follows from (\ref{C4}), in the pure limit we get $\sigma ^a_{xy}+\sigma ^b_{xy}=0$.
It should be noted that each of the terms $\sigma ^{b}_{xy}$ and
$\sigma ^{a}_{xy}$ is nonzero but they exactly cancel each other.
For example, calculating $\sigma ^{a}_{xy}$ we find
$\sigma ^{a}_{xy}=e^2\Delta /4\pi \epsilon _F$.

In the limit of $\omega \ll \Gamma _+$, we find from (\ref{C4})
\begin{equation}
\label{C5}
\sigma _{xy}^a+\sigma _{xy}^b=\frac{e^2}{4\pi }
\left[ \frac{\Delta }{\epsilon _F}
-\frac{b}{v}\frac{\Gamma _1}{\Gamma _+}
-\frac{b}{v}\frac{\Gamma _0}{\Gamma _+}\frac{\Delta}{\epsilon _F}
+\frac{c}{v}\frac{\epsilon _F}{\Gamma _+}
\left( 1-\frac{\Delta ^2}{\epsilon _F^2}\right)
\right]
\end{equation}
where we used the notation $\Upsilon _x=b\sigma _x+c\sigma _y$.
After substituting the values of $b$, $c$, and $\Gamma $
we find
\begin{equation}
\label{C6}
\sigma _{xy}^a+\sigma _{xy}^b=\frac{e^2\cos \theta}{4\pi }
\left[ 1-\frac{8\, (1+\cos ^2\theta )}
{(1+3\cos ^2\theta )^2}
\right]
\end{equation}
where $\cos \theta =\Delta /\epsilon _F$.
Calculating integral (\ref{C3}) we find
\begin{equation}
\label{C7}
\sigma _{xy}^{int}=-\frac{e^2\cos \theta }{4\pi }
\end{equation}
Combining it with (\ref{C5}) we obtain finally
\begin{equation}
\label{C8}
\sigma _{xy}=\sigma _{xy}^a+\sigma _{xy}^b+\sigma _{xy}^{int}
=-\frac{2e^2\cos \theta \, (1+\cos ^2\theta )}
{\pi \, (1+3\cos ^2\theta )^2}
\end{equation}
which coincides with Eq.(\ref{sxy2}) for $V_1=0$.

\section{Calculation of $\sigma^{II}_{xy}$}
\label{sigmaII_app}
In calculating $\sigma_{xy}^{II}$ it is useful to consider separately integrations over positive and negative $\epsilon$, i.e.
$\sigma_{xy}^{II}=I_{<}+I_{>}$, where with our simplifications
\begin{equation}
\begin{array}{l}
I_{<}=\frac{e^2}{4\pi} \int_{-\infty}^{0} d\epsilon f(\epsilon)
\int \frac{d^2 {\bf k}}{(2\pi)^2} (v_x^{+-}v_y^{-+}-\\
\\
-v_x^{-+}v_y^{+-}) (  \frac{1}   {\epsilon-\epsilon^{+}_{{\bf k}} } \frac{d}{d\epsilon}
(G^{A-}(\epsilon)- \\
\\
-G^{R-}(\epsilon)) + \frac{1}{(\epsilon^{+}_{{\bf k}}-\epsilon^{-}_{{\bf k}})^2}
(G^{A-}(\epsilon)-G^{R-}(\epsilon)) ).
\end{array}
\label{Iless1}
\end{equation}
We can get rid of derivative over $\epsilon$ by integration by parts and noticing that for $\epsilon<0$ and positive chemical potential
follows that $df(\epsilon)/d\epsilon=0$. Thus
\begin{equation}
\begin{array}{l}
I_{<}=e^2 \int_{-\infty}^{0} d\epsilon f(\epsilon)
\int \frac{d^2 {\bf k}}{(2\pi)^2}  \left(  \frac{i(v_x^{+-}v_y^{-+}-v_x^{-+}v_y^{+-})}{(\epsilon^{-}_{{\bf k}}-\epsilon^{+}_{{\bf k}})^2}
\delta(\epsilon -\epsilon_{{\bf k}}^{-})
 \right)
\end{array}
\label{Iless2}
\end{equation}
Further simplifications come from the fact that off-diagonal elements of the velocity operator are related to interband matrix elements of the
coordinate operator
$v^{+-}_x = (-i[\hat{x},H])^{+-}=iA_x^{+-} (\epsilon^{+}_{{\bf k_F}}-\epsilon^{-}_{{\bf k}})$ where
 $A_x^{+-} = \langle u^{+}_{{\bf k}} | i\partial/\partial k_x | u^{-}_{{\bf k}} \rangle $ etc.
Such elements have the property $i[A_x^{\pm \mp}A_y^{\mp \pm}-A_y^{\pm \mp}A_x^{\mp \pm}]=F^{\pm}$, where $F^+/F^-$
is the Berry curvature of the upper/lower sub-band. After this

\begin{equation}
I_{<}=-e^2
\int \frac{d^2 {\bf k}}{(2\pi)^2} F^{-}(k)
\label{Iless3}
\end{equation}
where we used the fact that the lower band is filled and hence $f(\epsilon_{{\bf k}}^{-})=1$.

Calculations of $I_{>}$ is analogous,
\begin{equation}
\begin{array}{l}
I_{>}=-\frac{e^2}{4\pi} \int_{0}^{+\infty} d\epsilon f(\epsilon)
\int \frac{d^2 {\bf k}}{(2\pi)^2} (v_x^{+-}v_y^{-+}- \\
\\
-v_x^{-+}v_y^{+-}) \Large{(}  \frac{1}   {\epsilon-\epsilon^{-}_{{\bf k}} } \frac{d}{d\epsilon}
(G^{A+}(\epsilon)-G^{R+}(\epsilon)) + \\
\\
+\frac{1}{(\epsilon^{-}_{{\bf k}}-\epsilon^{+}_{{\bf k}})^2}
(G^{A+}(\epsilon)-G^{R+}(\epsilon)) \Large{)}
\end{array}
\label{Ibigb}
\end{equation}
however for positive $\epsilon$ follows that $df(\epsilon)/d\epsilon=-\delta(\epsilon -\epsilon_F) \ne 0$ and thus
extra terms will appear after integration by parts.
\begin{equation}
I_{>}=I_{>}^{\prime} +I_{>}^{\prime \prime}
\label{Ibig1}
\end{equation}
where
\begin{equation}
I_{>}^{\prime}=-e^2
\int_{\epsilon_{{\bf k}}^{+}<\epsilon_F} \frac{d^2 {\bf k}}{(2\pi)^2}  F^{+}(k)
\label{Ibig2}
\end{equation}
and
\begin{equation}
\begin{array}{l}
I_{>}^{\prime \prime} =\frac{e^2} {4\pi} \int_{-\infty}^{+\infty} d\epsilon \frac{df(\epsilon)} {d\epsilon}
\int \frac{d^2 {\bf k} } {(2\pi)^2} \Large{[} \\
\\
 -\frac{ v_x^{+-} v_y^{-+}-v_x^{-+} v_y^{+-} } {  \epsilon^{+}_{{\bf k}}- \epsilon^{-}_{{\bf k}}   }
 (G^{R+}(\epsilon)-G^{A+}(\epsilon))  \Large{]}
\end{array}
\label{Ibig3}
\end{equation}
Comparing (\ref{Ibig3}) and (\ref{sxyint}) we find that they differ only by sign thus we can write that

\begin{equation}
\begin{array}{l}
\sigma^{II}_{xy} = -e^2\Large{(} \int_{\epsilon_{{\bf k}}^{+}<\epsilon_F} \frac{d^2 {\bf k}}{(2\pi)^2}  F^{+}(k) +\\
\\
+\int \frac{d^2 {\bf k}}{(2\pi)^2}  F^{-}(k)  \Large{)} - \sigma^{I(int)}_{xy}
\end{array}
\label{sigmaIIfin}
\end{equation}
We conclude that $\sigma^{II}_{xy}$ contains a term that exactly cancels the intrinsic term  $\sigma^{I(int)}_{xy}$ from $\sigma^I_{xy}$.

\end{document}